\documentclass[openacc]{rsproca_new}%%%%where rsproca is the template name

\usepackage{graphicx}
\usepackage{epstopdf, epsfig}
\usepackage{newtxtext}
\usepackage{newtxmath}
\usepackage{hyperref}
\usepackage{color}
\usepackage{siunitx}
\usepackage[square, numbers]{natbib}
\usepackage{soul}

\newcommand{\dd}[2]{\frac{\mathrm{d} #1}{\mathrm{d} #2}}

\newcommand{\order}[1]{\mathcal{O}(#1)}
\newcommand{\red}[1]{{\color{red} #1}}
\newcommand{\blue}[1]{{\color{blue} #1}}

\newcommand{\rout}[1]{\red{\st{#1}}}
\renewcommand{\rout}[1]{{}} %uncomment to remove any routed text.
\renewcommand{\blue}[1]{{\textcolor{black}{#1}}} %uncomment to make blue text render black 
\renewcommand{\red}[1]{{}} %uncomment to remove any routed text.

\newcommand{\epsone}{\epsilon_{1}} %eps1 = E0 alpha/ Cd
\newcommand{\epstwo}{\epsilon_{2}} %eps2 = E0 alpha/St
\newcommand{\epsthree}{\epsilon_{3}} %eps3 = tau/(L/c)
\newcommand{\epsfour}{\epsilon_{4}} 
\newcommand{\Pb}{\textit{P}_B}  %\frac{L/c}{ \tau}\frac{S_l - S_u}{2S_l},
\newcommand{\lt}{\delta} %dimensionless thermocline length, lt/l0
\newcommand{\Pt}{\textit{P}_T}

\newcommand{\cone}{c_1}  %c1 = eps1 / delta ~ O(1)
\newcommand{\ctwo}{c_2}  %c2 = eps2 / delta ~ O(1)
\renewcommand{\in}{\text{in}} %subcript on into and out of pycnocline variables
\newcommand{\out}{\text{out}}

\newcommand\abcdeqn[2]{\refstepcounter{equation}
     \[
     \label{#1}
     #2
     \eqno{\text{(\theequation)}\text{a,b,c,d}}
     \]
}

\newcommand\abeqn[2]{\refstepcounter{equation}
     \[
     \label{#1}
     #2
     \eqno{\text{(\theequation)}\text{a,b}}
     \]
}

\begin{document}

%%%% Article title to be placed here
\title{Asymptotic Analysis of Subglacial Plumes in Stratified Environments}

\author{%%%% Author details
Alexander T. Bradley$^{1}$, C. Rosie Williams$^{1}$, Adrian Jenkins$^{2}$ and Robert Arthern$^{1}$}

%%%%%%%%% Insert author address here
\address{$^{1}$British Antarctic Survey, Cambridge, UK\\
$^{2}$Department of Geography and Environmental Sciences, Northumbria University, Newcastle upon Tyne, UK.}

%%%% Subject entries to be placed here %%%%
\subject{Oceanography, Glaciology, Mathematical Modelling}

%%%% Keyword entries to be placed here %%%%
\keywords{Plumes, Antarctica, Ice-shelves, Ice-sheets, Melting}

%%%% Insert corresponding author and its email address}
\corres{Alexander T Bradley\\
\email{aleey@bas.ac.uk}}

%%%% Abstract text to be placed here %%%%%%%%%%%%
\begin{abstract}
Accurate predictions of basal melt rates on ice shelves are necessary for precise projections of the future behaviour of ice sheets. The computational expense associated with completely resolving the cavity circulation using an ocean model makes this approach unfeasible for multi-century simulations, and parametrizations of melt rates are required. At present, some of the most advanced melt rate parametrizations are based on a one-dimensional approximation to the melt rate that emerges from the theory of subglacial plumes applied to ice shelves with constant basal slopes and uniform ambient ocean conditions; in this work, we present an asymptotic analysis of the corresponding equations in which non-constant basal slopes and typical ambient conditions are imposed. This analysis exploits the small aspect ratio of ice shelf bases, the relatively weak thermal driving and the relative slenderness of the region separating warm, salty water at depth and cold, fresh water at the surface in the ambient ocean. We construct an approximation to the melt rate that is based on this analysis, which shows good agreement with numerical solutions in a wide variety of cases, suggesting a path towards improved predictions of basal melt rates in ice-sheet models.
\end{abstract}
%%%%%%%%%%%%%%%%%%%%%%%%%%%

%%%%%%%%%% Insert the texts which can accomdate on firstpage in the tag "fmtext" %%%%%

\begin{fmtext}

\end{fmtext}
\maketitle
\section{Introduction}\label{S:Introduction}
The Antarctic ice sheet has undergone significant change in recent years, characterized by pronounced thinning, mass loss and acceleration~\cite[e.g.][]{Pritchard2012Nature, Mouginot2014GRL}. An increase in basal melting on ice shelves -- the floating, tongue-like extensions of the grounded ice -- has been implicated in these changes~\cite{Shepherd2004GRL}; increased basal melting reduces the resistive effect that ice shelves apply to the grounded ice upstream~\cite{Gagliardini2010GRL} (commonly referred to as `buttressing') leading to an increase \blue{in} ice flux from the grounded to floating regions of the ice sheet~\cite{Gudmundsson2013Cryo}. Reductions in buttressing are thought to increase the possibility of ice sheet instabilities~\cite{Schoof2007JGeophysResEarth}. Changes in ice shelf buttressing and ice sheet instabilities are leading order controls on future projections of the Antarctic ice sheet and the associated sea level rise contributions~\cite{Arthern2017GRL}. Accurate descriptions of the basal melt rate that results from ice-ocean interactions are therefore necessary to constrain predictions of sea level rise. 

Basal melting on ice shelves is often characterized by three main modes, whose relative importance depends on myriad factors including the geometry of the ice shelf cavity, the presence of \rout{circumpolar deep water}\blue{Circumpolar Deep Water (CDW)} on the continental shelf, and oceanic tides~\cite{Jacobs1992JGlac}. So-called `mode one' melting is driven by inflows of cold, dense shelf water which is produced by sea ice formation, and results in the formation of buoyant plumes of meltwater; `mode two' melting is driven by warm \rout{Circumpolar Deep Water}\blue{CDW} that spills over the continental shelf; and `mode three' is driven by surface waters~\cite{Silvano2016Oceanography}. Mode one melting makes up the dominant contribution to melting of Antarctic ice shelves~\cite{Adusumilli2020NatureGeo}.

Ocean general circulation models aim to account for each of these modes of melting by explicitly resolving the circulation within ice shelf cavities. Such models require high resolution in both the horizontal and vertical directions if they are to accurately resolve important physical processes, such as the turbulent entrainment inside a boundary layer next to the ice~\cite[e.g.][]{Kimura2014JPhysOcean}. As a result, they are computationally expensive, especially when coupled to a dynamic ice sheet model whose geometry evolves over time; fully-coupled simulations of ice-ocean interactions are, to date, limited in their spatial extent to single ice shelf cavities~\cite[e.g.][]{DeRydt2016JGeophysResEarthSurf, Seroussi2017GRL} or in their temporal extent if multiple cavities are included~\cite[e.g.][]{Naughten2018JClim}.

An alternative approach is therefore required for multi-century simulations of ice sheets. As it stands, all simulations of the Antarctic ice sheets on this timescale have applied an analytic approximation -- or parametrization -- to the basal melt rate. Many different parametrizations have been proposed, but each typically falls into one (or more) of the following categories: parametrizations including simple ice draft depth dependence~\cite[e.g.][]{Joughin2014Science}; parametrizations based on a local balance of heat at the ice-ocean interface~\cite[e.g][]{Golledge2015Nature}; and parametrizations including \rout{with }a linear~\cite[e.g.][]{Favier2016Cryo} or quadratic dependence~\cite[e.g.][]{Holland2008JClimate} on ocean temperature.

In recent years, however, two novel melt rate parametrizations have emerged, which do not fall into these categories. The first, \blue{ described by Reese et al}.~\cite{Reese2018Cryo}, is based on a one-dimensional ocean-box model that coarsely resolves ice shelf cavities. The second is a `plume emulator' from \blue{Lazeroms et al.}~\cite{Lazeroms2018TheCryo}, which is based on the theory of subglacial plumes. Both of these conceptualize the ocean circulation in the ice shelf cavity, a feature that is lacking in the simpler parametrizations mentioned above. In this paper, we focus on the plume emulator parametrization, which produces patterns of melt rate that are qualitatively similar to observations when applied to Antarctic ice shelves~\cite{Lazeroms2018TheCryo} and performs reasonably well when compared to a fully coupled ice-ocean model~\cite{Favier2019GeosciModDev}.

The theory of subglacial plumes was pioneered by MacAyeal~\cite{MacAyeal1985} and Jenkins~\cite{Jenkins1991JGeophysResOceans}, studies which themselves build upon the classical theories of plumes and gravity currents~\cite[e.g.][]{Ellison1959JFM, Morton1956PRSL}. Subglacial plumes are buoyancy driven flows that form when freshwater enters the ocean adjacent to an ice shelf, by either melting or direct discharge at the grounding line; the freshwater is buoyant, resulting in a turbulent flow that moves upwards along the ice shelf base. The turbulent motion within the plume results in entrainment of relatively warm and salty ambient water; on the one hand, the associated increase in temperature of the plume promotes enhanced melting of the ice shelf base, further freshening the plume and increasing its buoyancy relative to the ambient ocean (ocean temperatures in Antarctica are typically within a few degrees of the local freezing point, so density variations are controlled primarily by salinity~\cite{Hewitt2020AnnRevFlu}). On the other hand, entrainment of ambient water results in an increase in the plume's salinity, thus reducing its buoyancy, and ultimately retarding its motion up the ice shelf base. This interplay between melting, ambient entrainment, and subglacial discharge leads to complex dynamics, which are further influenced by the variation of the pressure-dependent freezing temperature with depth. 

There are four important length scales involved in the dynamics of subglacial plumes. The first is related to stratification of the ambient ocean. Subglacial plumes entrain relatively salty (and thus dense) ambient water at depth. This water  is carried upwards to regions of lower ambient density, where it may ultimately become negatively buoyant; the depth at which this transition to negative buoyancy occurs prescribes a length scale associated with ambient stratification~\cite{Magorrian2016JGeoResOcean}. The second length scale is related to the variation of the freezing temperature with depth: as the plume rises, as does the in situ freezing point; the plume may become super-cooled, leading to refreezing of ice onto the shelf base. The distance over which the temperature difference between the plume and the in situ freezing point at the grounding line decays to zero as a result of this depth dependence on temperature gives the associated length scale~\cite{LaneSerff1995JGeophysResOceans}. The third length scale is associated with subglacial discharge. Far from the grounding line, mass inputs to the plume are dominated by ambient entrainment, but close to the grounding line, subglacial discharge may be important; the cross-over between these two regimes defines the length scale associated with subglacial discharge, which depends sensitively on the freshwater flux at the grounding line~\cite{Jenkins2011JPhysOcean}; ice shelves in Antarctica typically have a relatively small subglacial discharge, so the associated length scale is small in comparison with the length of the ice shelf and the effect of subglacial discharge on plume dynamics can typically be neglected. The final length scale is the plume thickness at which rotation (Coriolis forces) begin to dominate the dynamics. In practice, this is when the plume thickness exceeds the Ekman length~\cite{Jenkins2011JPhysOcean}. As the plume thickness approaches the Ekman length, the concept of entrainment becomes increasing poorly defined and simple plume theory becomes invalid.  Well-mixed conditions are likely to be maintained only within the Ekman layer, beyond which there will be a gradual transition to ambient properties through a broad pycnocline~\cite{Jenkins2021JPO}.  In a stratified ambient ocean, the distinction between ambient and plume will become unclear, and lower parts of the pycnocline will progressively detrain into the ambient.

The plume emulator parametrization of Lazeroms et al.~\cite{Lazeroms2018TheCryo} is based on an approximation to the melt rate that emerges from the system of equations describing subglacial plumes in one spatial dimension, which is then extended to two spatial dimensions. Originally, this approximation was based on a polynomial fit to data obtained by numerically solving this system of equations~\cite{Jenkins2014scaling}, but an analytic expression was subsequently derived by Lazeroms et al.~\cite{Lazeroms2019JPhysOcean}, via an asymptotic analysis of the system in the limit of weak thermal forcing and shallow slopes of the ice shelf base. This system of equations \rout{are }\blue{is} appropriate only for ice shelves with a constant basal slope, and whose adjacent ocean has a constant temperature and salinity; in practice, the ice shelf basal slope displays significant heterogeneity and the ambient ocean is often highly stratified. The aim of this paper is to construct an approximation to the melt rate that emerges from the equations describing a one-dimensional subglacial plume model which includes a typical ambient stratification and a non-constant ice shelf basal slope, with a view to ultimately improving upon the Lazeroms et al.~\cite{Lazeroms2018TheCryo} melt rate parametrization by accounting for more realistic conditions.

The ambient stratification we consider here is based on typical conditions observed in the Amundsen Sea sector of Antarctica~\cite{Jenkins2018NatureGeo}. It consists of two fairly uniform layers separated by a sharp pycnocline -- a relatively thin layer in which gradients in both temperature and salinity are strong. Below this pycnocline, the ambient ocean is warm and salty, deriving from modified \rout{Circumpolar Deep Water}\blue{CDW}, which spills over the continental shelf break. Above this pycnocline, the ambient ocean is cold and fresh, deriving from Winter Water which forms from cooling, brine drainage and convection beneath growing sea ice.

This paper is structured as follows: in \S\ref{S:Model}, we provide a brief outline of the mathematical model of subglacial plumes of~\cite{Jenkins1991JGeophysResOceans} and~\cite{Jenkins2011JPhysOcean}, in which ambient stratification and non-constant ice shelf basal slopes are included explicitly. This model is non-dimensionalized using scales that emerge from a similarity solution valid close to the grounding line. We then present example solutions of the equations governing this dimensionless model, which provide insight into the behaviour; in doing so, we informally introduce the different regions of the solution referred to in the asymptotic analysis that follows in \S\ref{S:Asymptotics}. This asymptotic analysis exploits the small size of several of the dimensionless parameters identified in the non-dimensionalization, including the aspect ratio of the ice shelf base, the weak thermal forcing, and the relative slenderness of the pycnocline referred to above. The behaviour is different in each of the four regions, which emerge naturally from the analysis, and we analyze each of these regions in turn. Following this, in \S\ref{S:MeltRate}, we present details of our approximation to the melt rate, which draws on the asymptotic analysis presented in \S\ref{S:Asymptotics}. In \S\ref{S:Numerics}, we assess the performance of this approximation by comparing it with both numerical solutions of the model equations and the corresponding Lazeroms et al.~\cite{Lazeroms2019JPhysOcean} approximation. Finally, in \S\ref{S:Discussion}, we discuss our results and provide concluding remarks.

\section{Mathematical Model}\label{S:Model}

\begin{figure}
\centering
\includegraphics[width = 0.9\textwidth]{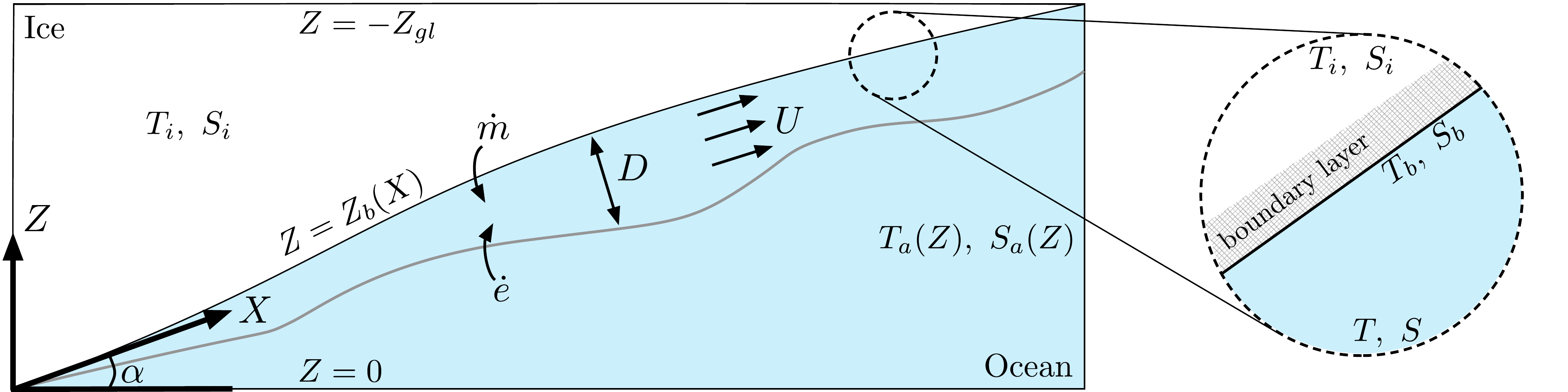}
\caption{Schematic diagram of a subglacial-plume. The plume originates at the grounding line, where the ice shelf base is inclined at an angle $\alpha$ to the horizontal. The grounding line is located at a depth $-Z_{gl} >0$ below sea level; we use shifted co-ordinates, with $Z = 0$ corresponding to the depth of the grounding line (i.e. sea level is at $Z = -Z_{gl}$). The plume is quasi-one-dimensional: it is described by its thickness $D(X)$, depth-averaged velocity $U(X)$, depth-averaged salinity $S(X)$, and depth-averaged temperature $T(X)$.  The plume gains mass by melting the ice shelf at a rate $\dot{m}$ and entraining ambient water at a rate $\dot{e}$. A boundary layer forms at the ice-ocean interface, across which heat and salt are exchanged; $T_b$ and $S_b$ denote the temperature and salinity on the ocean side of this boundary layer (see inset).}\label{fig:Schematic}
\end{figure}

We begin by providing a brief outline of the plume model described in~\cite{Jenkins1991JGeophysResOceans} and~\cite{Jenkins2011JPhysOcean}. The configuration is shown schematically in figure~\ref{fig:Schematic}. The $X$-axis is aligned with the ice-ocean interface (referred to henceforth as the ice shelf base) at the grounding line, and the $Z$-axis is aligned vertically; the origin of this co-ordinate system is the grounding line. The ice shelf base is located at $Z = Z_b(X)$, and the angle of the base at the grounding line is denoted by $\alpha$ [i.e. $\alpha = Z_b'(0)$, where $'$ denotes differentiation with respect to $X$].

Ultimately, we intend our results to  be applicable to ice shelves in Antarctica, which are characterized by ice shelf bases with low aspect ratios. \blue{The aspect ratio of ice shelves are usually largest at the grounding line, where they are typically lower than 1\%~\cite{Rignot2011GRL}; thus it is reasonable to assume that} $\alpha \ll 1$. In particular, this assumption means that the angle of inclination of the ice shelf base from the horizontal can be approximated by $Z_b'(X)$ (rather than a derivative with respect to an arc-length parameter measured along the ice-shelf base) and the distance from the grounding line along the ice shelf base can be approximated by the $X$ co-ordinate. The shallow slope assumption also means that the vertical depth and depth normal to the ice shelf base can be used interchangeably. We shall assume that the derivatives of $Z_b$ are well-defined everywhere, that the ice shelf base has a monotonic increasing profile, $Z_b'(X) >0$, and that \blue{the slope of the ice shelf base, although non-constant, remains on the order of its slope at the grounding line everywhere} \rout{$\alpha$ is a characteristic slope throughout} [i.e. $Z_b'(X)\sim \alpha$ for all $X >0$]. 

The plume is described in terms of its depth-averaged thickness $D(X)$, velocity $U(X)$, temperature $T(X)$, and salinity $S(X)$ (figure~\ref{fig:Schematic}). The depth-average plume density is denoted by $\rho(X)$. We neglect any time dependence, since the timescale on which the geometry of ice shelf base evolves is significantly longer than the timescale on which the plume reaches equilibrium with the present geometry~\cite{Hewitt2020AnnRevFlu}. \blue{We employ a Bousinnesq approximation, in which the plume and ambient are considered to have the same density $\rho_0$, except for in contributions to the buoyancy force, where the density difference is multiplied by gravitational acceleration, $g$. In addition, we assume that the specific heat capacities of both the ice and the ocean are constant, and denote these by $c_i$ and $c$, respectively.}

Temperature and salinity are exchanged across the ice-ocean interface; the temperature and salinity on the \rout{plume} \blue{ocean} side of this interface are denoted by $T_b$ and $S_b$, respectively (see figure~\ref{fig:Schematic}). A boundary layer forms within the ice at the ice-ocean interface; the temperature and salinity within in the ice, far from this boundary layer, are denoted by $T_i$ and $S_i$, respectively. The ambient ocean has salinity $S_a(Z)$ and temperature $T_a(Z)$; both of these feed into the ambient density $\rho_a(Z)$, through which the plume `feels' the stratification of the ambient ocean.  We shall discuss in due course our choice of ambient temperature and salinity profiles which  impose the aforementioned two layer stratification. 

\rout{The plume is separated from the ice by a molecular sublayer, across which temperature and salinity are exchanged (inset in figure~\ref{fig:Schematic}); the salinity and temperature on the plume sides of this boundary layer are denoted by $S_b$ and $T_b$ respectively. The ambient ocean has salinity $S_a(Z)$ and temperature $T_a(Z)$, both of which feed into the ambient density $\rho_a(Z)$, through which the plume `feels' the stratification of the ambient ocean.  We shall discuss in due course our choice of ambient temperature and salinity, which permit us to impose the aforementioned two layer stratification. }

\subsection{Conservation Equations} 
After applying the assumptions described above,\rout{ and making a standard Boussinesq approximation,} conservation of mass, momentum, salt, and heat require the following system of ordinary differential equations (ODEs) to hold~\cite{Jenkins2011JPhysOcean,Magorrian2016JGeoResOcean}:
\begin{align}
\dd{(DU)}{X} &= \dot{e} + \dot{m},\label{E:Modelling:Model:RawMass}\\
\dd{(DU^2)}{X} &= \frac{D\Delta \rho}{\rho_0} g \dd{Z_b}{X}  -\red{\frac{\tau_T}{\rho_0}} \blue{C_d U^2},\label{E:Modelling:Model:RawMometum}\\
\dd{(DUS)}{X} &= \dot{e}S_a + \dot{m}S_b -\frac{Q_S}{\rho_0}\label{E:Modelling:Model:RawSalt}\\
\dd{(DUT)}{X} &= \dot{e}T_a + \dot{m}T_b -\frac{Q_T}{c \rho_0}.\label{E:Modelling:Model:RawTemp}
\end{align}
\blue{Note that in~\eqref{E:Modelling:Model:RawMometum}, we have assumed that the flux of momentum from the plume to the ice shelf base follows a quadratic drag law with a drag coefficient $C_d$, as is standard when modelling of subglacial plumes.} Here\rout{ $\tau_T$,} $Q_s$ and $Q_T$ are the turbulent fluxes of\rout{ momentum,} heat\rout{,} and salt to the ice shelf base. We take the following parametrizations for these quantities, based on a\rout{ quadratic drag law and} shear dominated viscous sublayer:
\begin{align}
Q_S &= \rho_0 \mathrm{St}_S U (S - S_b),\label{E:Modelling:Model:TurbulentSalt} \\
Q_T &= c \rho_0 \mathrm{St}_T U (T -T_b)\red{,}\blue{.}\label{E:Modelling:Model:TurbulentHeat}
\end{align}
Here $\mathrm{St}_S, \mathrm{St}_T$ are (constant) haline and thermal Stanton numbers, respectively, which parametrize exchange across a sublayer that forms on the ocean side of the ice-ocean interface\rout{, $C_d$ is a drag coefficient, and $c_i$ is the specific heat capacity of ice}. Typical values of all constants introduced here are listed in table~\ref{T:Constants}.

In~\eqref{E:Modelling:Model:RawMass}--\eqref{E:Modelling:Model:RawTemp}, $\dot{m}$ is the melt rate and $\dot{e}$ is the entrainment rate of ambient water (figure~\ref{fig:Schematic}); here we take a standard entrainment parametrization, 
\begin{equation}\label{E:Modelling:Model:Entrainment}
\dot{e} = E_0 U \dd{Z_b}{X},
\end{equation}
which is based on a competition between shear-induced turbulence and the stabilizing effect of mixing across a sharp density interface~\cite{Pedersen1980,Turner1986JFM}. The parametrization~\eqref{E:Modelling:Model:Entrainment} suggests that the entrainment coefficient $\dot{e}/U = E_0 Z_b'(X)$\blue{, which is consistent with simple field observations~\cite{Pedersen1980} and laboratory experiments~\cite{Wells2005GAF}. We use this parametrization for simplicity, but note that the exact form of appropriate entrainment coefficients remains the subject of significant enquiry; in particular, several studies have suggested more complex Richardson number dependent parametrizations~\cite[e.g.][]{Holland2007JGeophysResOceans}, but these have been shown in related studies of plume dynamics to yield similar results to the simple parameterization employed here}~\cite{Magorrian2016JGeoResOcean}.\rout{has no Richardson number dependence; we use this parametrization for simplicity, but note that entrainment is typically Richardson number dependent.}

In the momentum equation~\eqref{E:Modelling:Model:RawMometum}, we have introduced the buoyancy deficit $\Delta \rho$. We take a linear equation of state,
\begin{equation}\label{E:Modelling:Model:EquationOfState}
\frac{\Delta \rho}{\rho_0}  = \frac{\rho_a - \rho}{\rho_0}= \beta_S(S_a - S) - \beta_T (T_a - T).
\end{equation}
Here $\rho_0$ is a reference density, $\beta_S$ is a haline contraction coefficient, and $\beta_T$ is a thermal expansion coefficient. 

Note that in~\eqref{E:Modelling:Model:RawMometum} we have neglected a term arising from the difference between the slope of the plume-ambient interface and the slope of the ice shelf base, as is standard when describing subglacial plumes~\cite{Hewitt2020AnnRevFlu}. This term only becomes important when the plume thickness is comparable to the length scale associated with freezing point dependence on depth~\cite{Jenkins2011JPhysOcean}. Since this thickness is typically beyond the Ekman length, plume theory is not expected to be applicable here (see \S\ref{S:Introduction}).

\begin{table}[!h]
\caption{Typical parameter values used in the plume model. Values in the top half are from~\cite{Jenkins1991JGeophysResOceans}, with the exception of the drag coefficient $C_d$ and entrainment coefficient $E_0$, where we list only the order of magnitude to reflect the uncertainty in these values~\cite{Hewitt2020AnnRevFlu}. The values towards the bottom, relating to the pycnocline, are interpreted from~\cite{Jenkins2018NatureGeo}, and reflect typical conditions for the Amundsen Sea Sector of Antarctica.}\label{T:Constants}
\begin{center}
\begin{tabular}{llll}%%%The number of columns has to be defined here
\hline
Symbol & Description & Typical Value & Units \\
\hline
$\alpha$ &Grounding line basal slope &$3\times 10^{-3}$ &- \\
$E_0$   & Entrainment coefficient    & $10^{-2}$ & -       \\
$\Gamma$      & Freezing point salinity coefficient & $5.73 \times 10^{-2}	$    	  &  $\si{\celsius}~\text{PSU}^{-1}$     \\
$T_o$      & Freezing point offset & $8.32 \times 10^{-2}	$    	  &  $\si{ \celsius}$     \\
$\lambda$      & Freezing point depth coefficient & $7.61 \times 10^{-4}	$    	  &  $\si{\celsius . \meter^{-1}}$     \\
$C_d$ & Drag coefficient & $10^{-3}$ & - \\
$\mathrm{St}$ & Composite Stanton number & $5.9 \times 10^{-4}$ & -\\
$\beta_S$ & Haline contraction coefficient & $7.86 \times 10^{-4} $ & $\text{PSU}^{-1}$ \\
$\beta_T$ & Thermal expansion coefficient & $3.87\times 10^{-5}$ & $\si{\celsius^{-1}}$ \\
$L$ & Latent heat of fusion of ice & $3.35 \times 10^5 $& $\si{\joule . \kg^{-1}}$\\
$c$ & Specific heat capacity ocean water & $3.974 \times 10^3$ & $\si{\joule .\kilogram .\celsius^{-1}}$\\
$c_i$ & Specific heat capacity ice & $2.009 \times 10^3$ & $\si{\joule . \kilogram .\celsius^{-1}}$\\

$\rho_0$ &  Reference ocean density & $1000$ & $\si{\kilogram . \meter^{-3}}$\\
$T_{l}$ & Ambient temperature lower layer &$0.5$        &  $\si{ \celsius}$   \\
$T_{u}$ & Ambient temperature upper layer &$-1.5$        &  $\si{ \celsius}$  \\
$S_{l}$ & Ambient salinity lower layer &$34.6$        &  $\text{PSU}$   \\
$S_{u}$ & Ambient salinity upper layer &$34.0$        &  $\text{PSU}$  \\
$\ell_p$ & Half lengthscale of pycnocline & $50$ & $\si{\meter}$\\
$Z_{gl}$ & Negative grounding line depth & $-1500$ & $\si{\meter}$\\

\hline
\end{tabular}
\end{center}
\vspace*{-4pt}
\end{table}

\subsubsection{Melting}
To close the system of equations~\eqref{E:Modelling:Model:RawMass}--\eqref{E:Modelling:Model:EquationOfState}, we must specify the melt rate $\dot{m}$.  Often, the so-called `three-equation formulation' is applied as such a closure; in this formulation, the ice-water interface is assumed to be in local thermodynamic equilibrium, so its temperature and salinity satisfy the liquidus condition,
\begin{equation}
    T_b = T_o + \lambda (Z_{b} + Z_{gl}) - \Gamma S_b,\label{E:Modelling:Model:Closure:3eqn3}
\end{equation}
which relates the ice-ocean interface temperature to the local pressure freezing point~\cite{Holland1999JPhysOcean} [note that $-(Z_b + Z_{gl})$ is the depth below sea level]. It is through the liquidus condition~\eqref{E:Modelling:Model:Closure:3eqn3} that the plume `knows' about depth dependence of the freezing point. Heat and salt balances at the interface require the following to hold:
\begin{align}
\dot{m}L + \dot{m}c_i (T_b - T_i) &= c \mathrm{St}_T U(T - T_b),\label{E:Modelling:Model:Closure:3eqn1}\\
\dot{m}(S_b - S_i) &= \mathrm{St}_S U(S - S_b).\label{E:Modelling:Model:Closure:3eqn2}
\end{align}  
Here\rout{ $c, c_i$ are the specific heat capacities of water and ice, respectively,} $L$ is the latent heat of fusion, $\Gamma$ is the freezing point salinity coefficient,  $T_o$ is a freezing point offset, and $\lambda$ is the freezing point depth coefficient. The terms on left-hand side of equation~\eqref{E:Modelling:Model:Closure:3eqn1} relate to phase change and heat conduction, respectively, with the latter term based on steady state one-dimensional advection-diffusion and is valid for P\'{e}clet numbers\rout{ less} \blue{greater} than five~\cite{Holland1999JPhysOcean} \blue{[a typical P\'{e}clet number, based on an ice shelf of thickness 1000~m melting at $\order{10}$~m~$\text{year}^{-1}$, is on the order of $100$~\cite{Holland1999JPhysOcean}].}

A common alternative to the melting closure~\eqref{E:Modelling:Model:Closure:3eqn3}--\eqref{E:Modelling:Model:Closure:3eqn2} is the `two-equation formulation'~\cite{McPhee1992JGeophysResOcean}. This formulation uses the following approximation to the heat balance~\eqref{E:Modelling:Model:Closure:3eqn2}:
\begin{equation}
\dot{m}L + \dot{m}c_i (T_f - T_i) = c~ \mathrm{St} ~U(T - T_f),\label{E:Modelling:Model:Closure:2eqn1}
%T_f &= T_o + \lambda (Z_{gl} + Z_b) - \Gamma S.\label{E:Modelling:Model:Closure:2eqn2}
\end{equation}
Here, $T_f =T_o + \lambda (Z_{gl} + Z_b) - \Gamma S$ is the freezing temperature outside the boundary layer on the underside of the ice shelf, and $\mathrm{St}$ is a composite Stanton number, which parametrizes exchanges of heat and salt across the boundary layer simultaneously.

As in Jenkins~\cite{Jenkins2011JPhysOcean}, we use the three equation formulation~\eqref{E:Modelling:Model:Closure:3eqn3}--\eqref{E:Modelling:Model:Closure:3eqn2} to eliminate $(T-T_b)$ and $(S-S_b)$ from the turbulent flux equations~\eqref{E:Modelling:Model:TurbulentSalt} and~\eqref{E:Modelling:Model:TurbulentHeat}, and then use the two equation formulation subsequently. \blue{Essentially,  Jenkins~\cite{Jenkins2011JPhysOcean} treats the three-equation formulation and the two-equation formulation as interchangeable; this is justified therein on the basis that the formulations work equally well in several observational studies~\cite{Jenkins2010JPO}. We note that, in addition, similar studies to that considered here have suggested only a small difference (maximum 2\%) between solutions using the two- and three-equation formulations, provided that the ice shelf base remains relatively shallow, as we do here~\cite{Lazeroms2019JPhysOcean}}.  From~\eqref{E:Modelling:Model:Closure:2eqn1}, the melt rate is then
\begin{equation}\label{E:Modelling:Model:Closure:mdot}
\dot{m} = M_0 U \Delta T,  
\end{equation}
where 
\abeqn{E:Modelling:Model:MeltRatePrefAndEffTemp}{
M_0 = \frac{\mathrm{St}}{T_f - T_i^{\text{ef}}}, \qquad T_i^{\text{ef}} = T_f - \frac{L + c_i(T_f - T_i)}{c}}
are the melt rate prefactor and effective temperature of melt water from the ice, respectively, and $\Delta T = T - T_f$ is the thermal driving. Note that while our derivation of~\eqref{E:Modelling:Model:Closure:mdot}--\eqref{E:Modelling:Model:MeltRatePrefAndEffTemp} does not preclude the possibility of negative melt rates (correspond to refreezing of ice), we have not included any description frazil ice crystal growth in our model~\cite{Jenkins1995}. Including frazil ice growth can significantly alter the behaviour of subglacial plumes in regions where the melt rate is negative~\cite{ReesJones2018Cryo}; its lack of representation is a shortcoming of the model presented here.

The freezing and internal temperatures of the ice are typically within a few degrees of one another, so $c_i(T_f - T_i)/ L \ll 1$, and thus the effective temperature of released melt water is dominated by latent heat: $T_i^{\text{ef}} \approx -L/c \approx -84\si{\celsius}$. It is therefore reasonable to neglect the contribution to the melt rate from diffusive heat flux. In this case the melt rate prefactor~\eqref{E:Modelling:Model:MeltRatePrefAndEffTemp}a reduces to
\begin{equation}\label{E:Modelling:Model:Closure:M0_simple}
 M_0 = \frac{c~\mathrm{St}}{L}.
\end{equation}
Similarly, the effective temperature of melt water from the ice is significantly colder than ambient, so the effective melt water density contrast can also be simplified by making the following approximation:
\begin{equation}\label{E:Modelling:Model:EffectiveDensity}
\Delta \rho_i^{\text{ef}} \coloneqq \rho_0 \left[\beta_S\left(S_a - S_i\right) - \beta_T\left(T_a - T_i^{\text{ef}}\right)\right]  \approx \rho_0 \left[\beta_S\left(S_a - S_i\right) - \beta_T\frac{L}{c}\right].
\end{equation}
%at this point we have the equations we will work with henceforth (i.e. all the terms here will be retained)

\subsubsection{Reformulation in terms of buoyancy deficit and thermal driving}
\blue{To proceed, we first insert}\rout{Inserting} the expressions~\eqref{E:Modelling:Model:TurbulentSalt}--\eqref{E:Modelling:Model:TurbulentHeat} for the turbulent flux, the entrainment parametrization~\eqref{E:Modelling:Model:Entrainment}, the heat and salt balances~\eqref{E:Modelling:Model:Closure:3eqn1}--\eqref{E:Modelling:Model:Closure:3eqn2}, and the melt rate relation~\eqref{E:Modelling:Model:Closure:M0_simple} into the conservation equations~\eqref{E:Modelling:Model:RawMass}--\eqref{E:Modelling:Model:RawTemp}\blue{. Jenkins~\cite{Jenkins2011JPhysOcean} demonstrated that the resulitng}\rout{gives a} system of equations \rout{that }can be rearranged such that they are expressed as conservation equations for fluxes of mass, momentum, buoyancy deficit \blue{$\Delta \rho$}, and thermal driving \blue{$\Delta T$}; using the approximations~\eqref{E:Modelling:Model:Closure:M0_simple} and~\eqref{E:Modelling:Model:EffectiveDensity}, this system of equations reads
\begin{align}
\dd{(DU)}{X} &= E_0 U \dd{Z_b}{X}+ \frac{c~\mathrm{St}}{L}U\Delta T,\label{E:Modelling:Model:Mass}\\
\dd{(DU^2)}{X} &= g \dd{Z_b}{X} \frac{\Delta \rho }{\rho_0} D - C_d U^2, \label{E:Modelling:Model:Mom}\\
\dd{(DU\Delta \rho)}{X} &= \frac{c~\mathrm{St}~\rho_0}{L}\left(\beta_S S_a - \beta_T \frac{L}{c}\right) U\Delta T + \rho_0\left(\beta_S \dd{S_a}{Z}-\beta_T \dd{T_a}{Z}\right) \dd{Z_b}{X} DU, \label{E:Modelling:Model:Buoyancy}\\
\dd{(DU\Delta T)}{X} &= \left\{T_a - T_{f,gl} + \Gamma \left[S_a - S_a(0)\right] - \lambda Z_b\right\}E_0 U \dd{Z_b}{X} - \mathrm{St}U \Delta T - \lambda \dd{Z_b}{X} DU.
\label{E:Modelling:Model:Thermal}
\end{align}
Here we have set the ice salinity $S_i = 0$ and introduced
\begin{equation}
T_{f,gl} = T_o + \lambda Z_{gl} - \Gamma S_a(0),
\end{equation}
which is the freezing temperature associated with the ambient salinity at the depth of the grounding line.

\subsubsection{Ambient Stratification}\label{S:Model:Pycnocline}
\begin{figure}
\centering
\includegraphics[scale =0.4]{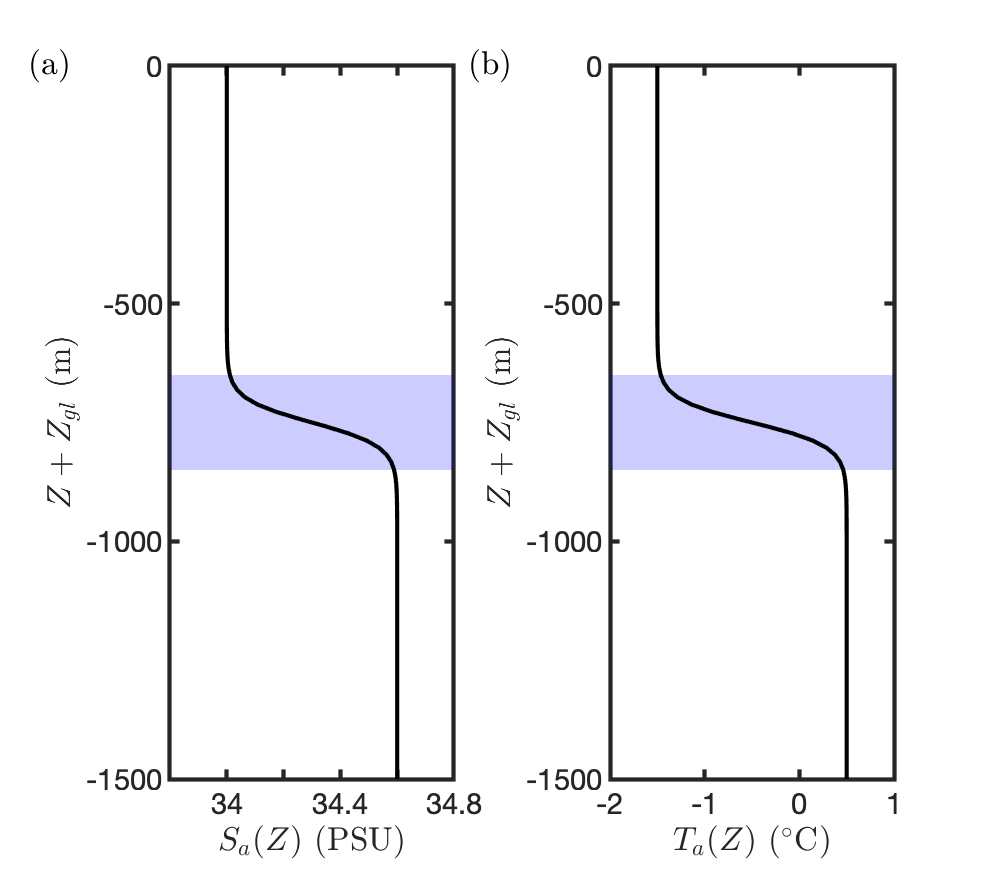}
\caption{Plot of (a) the ambient salinity profile and (b) the ambient temperature profile given by~\eqref{E:Modelling:Pycnocline:AmbientProfiles} as a function of \rout{distance above sea level,} $Z + Z_{gl}$, using a grounding line depth $Z_{gl} = -1500~\si{\meter}$ and pycnocline lengthscale $\ell_p = 50~\si{\meter}$. In both plots, the shaded area indicates the region $Z_p - 2\ell_p < Z < Z_p + 2\ell_p$ in which the majority of the change in the salinity and temperature occurs.}\label{fig:Pycnocline_Profiles}
\end{figure}

As discussed, we assume a two-layer ambient stratification in which warm, salty water of constant temperature $T_l$ and salinity $S_l$ in a lower layer sits beneath an upper layer of cold, fresh water of constant temperature $T_u < T_l$ and salinity $S_u < S_l$. The two layers are separated by a thin pycnocline centred at a depth $Z = Z_p$, which has a characteristic thickness scale $\ell_p$ across which the temperature and salinity vary. We impose this ambient stratification by taking
\begin{equation}\label{E:Modelling:Pycnocline:AmbientProfiles}
S_a = \frac{1}{2}\left[S_l + S_u - (S_l - S_u)\tanh\left(\frac{Z - Z_p}{\ell_p}\right)\right], \quad
T_a = \frac{1}{2}\left[T_l + T_u - (T_l - T_u)\tanh\left(\frac{Z - Z_p}{\ell_p}\right)\right],
\end{equation} % \label{E:Modelling:Pycnocline:Ta}
in~\eqref{E:Modelling:Model:Mass}--\eqref{E:Modelling:Model:Thermal}. Profiles of~\eqref{E:Modelling:Pycnocline:AmbientProfiles} are shown in figure~\ref{fig:Pycnocline_Profiles}. 

\subsection{Similarity Solution}\label{S:Model:SimilaritySolution}
Before we proceed with a non-dimensionalization of the model equations~\eqref{E:Modelling:Model:Mass}--\eqref{E:Modelling:Model:Thermal}, it is instructive to consider the behaviour near the grounding line. There, the in situ freezing point is approximately constant, the ice shelf base has an approximately constant slope [$Z_b(X) \approx \alpha X$], and the plume has no information about gradients in the ambient temperature or salinity. Under these conditions, the problem has no intrinsic length scale and a similarity solution to~\eqref{E:Modelling:Model:Mass}--\eqref{E:Modelling:Model:Thermal} exists~\cite{Magorrian2016JGeoResOcean}:
\begin{align}
D &= \frac{2}{3}E_0 \alpha X, & U &= \left[\frac{2E_0 \alpha^2 g}{\left(4E_0 \alpha + 3 C_d\right)} \frac{\Delta \rho}{\rho_0}\right]^{1/2}X^{1/2},\label{E:Similarity:SimilaritySol1}\\
 \frac{\Delta \rho}{\rho_0} &= \frac{\mathrm{St}}{E_0 \alpha}\frac{\Delta T}{L/c}\left(\beta_S S_l - \beta_T \frac{L}{c}\right), & \Delta T &= \frac{E_0 \alpha}{E_0 \alpha + \mathrm{St}}\tau.\label{E:Similarity:SimilaritySol2}
\end{align}
where $\tau = T_l - T_{f,gl}$ is the difference between the ambient and local freezing temperatures at the grounding line. Note that in~\eqref{E:Similarity:SimilaritySol1}--\eqref{E:Similarity:SimilaritySol2}, we have neglected the mass contribution from ice shelf melting\blue{, which is unimportant in comparison with the mass contribution from entrainment in proximity to the grounding line~\cite{Hewitt2020AnnRevFlu}}\rout{;} \blue{(we shall demonstrate this explicitly in \S\ref{S:Asymptotics}).} \rout{ that this is reasonable for locations close to the grounding line.}

The Richardson number, which quantifies the relative strength of buoyancy and drag, associated with~\eqref{E:Similarity:SimilaritySol1}--\eqref{E:Similarity:SimilaritySol2} is
\begin{equation}\label{E:Similarity:Richardson}
\mathrm{Ri} = \frac{\Delta \rho}{\rho_0}\frac{D g \cos \alpha}{U^2} = \frac{4E_0 \alpha + 3C_d}{3} \frac{\cos \alpha}{\alpha} \approx \frac{C_d}{\alpha},
\end{equation}
where the final approximation makes use of $E_0 \alpha \ll C_d$ and $\alpha \ll 1$ (table~\ref{T:Constants}). In general, if the Richardson number is less than unity, the plume velocity is greater than its internal wave speed and disturbances cannot propagate upstream; in this case numerical solutions of~\eqref{E:Modelling:Model:Mass}--\eqref{E:Modelling:Model:Thermal} are stable to perturbations~\cite{Jenkinsphdthesis}. If, however, the Richardson number is greater than unity, interfacial waves can propagate upstream, and solutions of~\eqref{E:Modelling:Model:Mass}--\eqref{E:Modelling:Model:Thermal} are numerically unstable. In practice, the Richardson number emerges from the solution of~\eqref{E:Modelling:Model:Mass}--\eqref{E:Modelling:Model:Thermal}, which must be determined numerically in general. However, if we take the local approximation~\eqref{E:Similarity:Richardson} as a proxy for the global Richardson number, we require ice shelf bases to be sufficiently steep, i.e. $C_d/\alpha < 1$, and thus restrict ourselves to ice shelf geometries satisfying this condition henceforth. \blue{We note that this is consistent with our earlier assumption that $\alpha \ll 1$, since the drag coefficient is also small, with $C_d < \alpha$ (table~\ref{T:Constants}). This restriction on the shallowness of ice shelf bases is made purely to ensure numerical stability of the full plume model equations, thus facilitating a comparison between the approximation constructed here and such numerical solutions; in particular, it is not a necessary condition in the construction of the approximation and is therefore not expected to impact on its applicability.}

\subsection{Non-dimensionalization}\label{S:Model:NonDim}
We use the similarity solution~\eqref{E:Similarity:SimilaritySol1}--\eqref{E:Similarity:SimilaritySol2} to provide relationships between the scales of the variables, which we denote by square brackets:
\begin{align}
\left[D\right] &= E_0 \alpha \left[X\right], & \left[U\right] &= \left(\frac{\beta_S S_l g E_0 \alpha^2 \tau}{C_d ~L/c}\right)^{1/2}[X]^{1/2},\label{E:Similarity:Scales1}\\
\left[\Delta \rho\right] &= \beta_S S_l \rho_0 \frac{\tau}{L/c}, & \left[\Delta T\right] &= \frac{E_0 \alpha}{\mathrm{St}}\tau.\label{E:Similarity:Scales2}
\end{align}
Note that variable scales have been simplified slightly from those expected from~\eqref{E:Similarity:SimilaritySol1}--\eqref{E:Similarity:SimilaritySol2} by using $E_0 \alpha \ll C_d$, $\mathrm{St}$ and $\beta_T L /c \ll \beta_S S_l$.

We take the vertical lengthscale $\left[Z\right] = \ell= \tau / \lambda$, which is the vertical distance associated with a change in the local freezing point from a temperature $\tau$ to zero degrees. The corresponding length scale in the $X$-direction is $\left[X\right] = \ell / \alpha=\tau/ (\alpha \lambda)$.

Inserting $\left[X\right]$ into~\eqref{E:Similarity:Scales1}--\eqref{E:Similarity:Scales2} gives a consistent set of variable scales; the problem is non-dimensionalized by introducing variables scaled accordingly (denoted by  $\hat{.}$):
\begin{align}
D&= E_0 \ell \hat{D}, & U&= \left(\frac{E_0 \alpha}{C_d} \frac{\tau}{L/c}~\beta_s S_l ~\ell  g\right)^{1/2}\hat{U}, & Z_b &= \ell \hat{Z}_b,\label{E:NonDim:NonDimVar1}\\
\frac{\Delta \rho}{\rho_0} &= \beta_S S_l \frac{\tau}{L/c}\Delta \hat{\rho}, & \Delta T &= \frac{E_0 \alpha}{St}\tau \Delta \hat{T}, & X &=\frac{\ell}{\alpha}\hat{X}. \label{E:NonDim:NonDimVar2}
\end{align}

Combining model equations~\eqref{E:Modelling:Model:Mass}--\eqref{E:Modelling:Model:Thermal} with the ambient salinity and temperature~\eqref{E:Modelling:Pycnocline:AmbientProfiles} and inserting~\eqref{E:NonDim:NonDimVar1}--\eqref{E:NonDim:NonDimVar2} gives
\begin{align}
\dd{(\hat{D}\hat{U})}{\hat{X}} &= \hat{U} \dd{\hat{Z}_b}{\hat{X}} +\epsthree \hat{U} \Delta \hat{T},\label{E:NonDim:mass}\\
\epsone \dd{(\hat{D}\hat{U}^2)}{\hat{X}} &= \hat{D} \Delta \hat{\rho} \dd{\hat{Z}_b}{\hat{X}} - \hat{U}^2,\label{E:NonDim:mom} \\
\dd{(\hat{D}\hat{U}\Delta \hat{\rho})}{\hat{X}}  &= -\frac{\Pb}{\delta} \mathrm{sech}^2\left(\frac{\hat{X} - \hat{X}_p}{\lt}\right)\dd{\hat{Z}_b}{\hat{X}} \hat{D}\hat{U}+
 \left[\kappa - \epsfour \tanh \left(\frac{\hat{X} - \hat{X}_p}{\lt}\right) \right] \hat{U} \Delta \hat{T}\label{E:NonDim:buoyancy} \\
\epstwo \dd{(\hat{D}\hat{U}\Delta \hat{T})}{\hat{X}} &= \left\{1 - \hat{Z}_b - \Pt\left[1 + \tanh\left(\frac{\hat{X} - \hat{X}_p}{\lt}\right)\right]\right\} \hat{U}\dd{\hat{Z}_b}{\hat{X}}  - \hat{U}\Delta \hat{T}- \hat{D}\hat{U}\dd{\hat{Z}_b}{\hat{X}}.\label{E:NonDim:thermal}
 \end{align}Here $\hat{X}_p$ is the $\hat{X}$ co-ordinate at which the ice shelf base passes through the centre of the pycnocline, i.e. $\hat{Z}_b(\hat{X}_p) = \hat{Z}_p$. We shall assume that the pycnocline is located away from the grounding line, which is encoded by taking $\hat{X}_p = \order{1}$.

%\begin{table}[!h]
%\caption{Dimensionless parameters in the model %equations~\eqref{E:NonDim:mass}--\eqref{E:NonDim:thermal}. Typical values are based on %the values listed in table~\ref{T:Constants}.}\label{T:Dimensionless_Parameters}
%\begin{center}
%\begin{tabular}{llll}
%Parameter name  & Definition   & Typical Value \\
%\hline 
%\vspace{2pt}
%\vspace{3pt}$\epsone$ & $  \frac{E_0 \alpha}{C_d} $  &$ 3 \times 10^{-2}$ \\
%\vspace{2pt}$\epstwo$ & $  \frac{E_0 \alpha}{St} $ & $ 5 \times 10^{-2}$ \\
%\vspace{2pt}$\epsthree$ & $  \frac{\tau}{L/c} $ & $ 4 \times 10^{-2}$ \\
%\vspace{2pt}$\epsfour$ & $  \frac{S_l - S_u}{2 S_l} $  &$ 2 \times 10^{-2}$ \\
%\vspace{2pt}$\lt$ & $  \frac{\ell_p}{\ell} $ & $  5 \times 10^{-3}$ \\
%\vspace{2pt}$\Pb$ & $  \frac{L/c}{ \tau}\frac{S_l - S_u}{2S_l}\left[1 - \frac{\beta_T %(T_l - T_u)}{\beta_S (S_l - S_u)}\right]$   & $  0.2$ \\
%\vspace{2pt}$\Pt$ & $ \frac{T_l - T_u + \Gamma( S_l - S_u)}{2 \tau}$ & $  1.2$ \\
%\vspace{2pt}$\kappa$ & $ \frac{S_l + S_u}{2S_l} -  \frac{\beta_T L/c}{ \beta_S S_l}$ &  %$  1.2$ \\
%\hline
%\vspace{2pt}$k_2$ & $\frac{\epstwo}{\epsone}$ & 1.7\\
%\vspace{2pt}$k_3$ & $\frac{\epsthree} {\epsone}$ & 1.4\\
%\vspace{2pt}$k_4$ & $\frac{\epsfour}{\epsone}$ & 0.3\\
%\end{tabular}
%\end{center}
%\vspace*{-4pt}
%\end{table}

\begin{table}[!h]
\caption{Dimensionless parameters in the model equations~\eqref{E:NonDim:mass}--\eqref{E:NonDim:thermal}. Typical values are based on the values listed in table~\ref{T:Constants}.}\label{T:Dimensionless_Parameters}
\begin{center}
\begin{tabular}{llll}
Parameter name  & Definition   & Typical Value \\
\hline 
\vspace{2pt}
$\epsone$ & $  E_0 \alpha /C_d $  &$ 3 \times 10^{-2}$ \\
$\epstwo$ & $  E_0 \alpha \blue{/}\mathrm{St} $ & $ 5 \times 10^{-2}$ \\
$\epsthree$ & $  \tau c / L $ & $ 4 \times 10^{-2}$ \\
$\epsfour$ & $  (S_l - S_u)/(2 S_l) $  &$ 2 \times 10^{-2}$ \\
$\lt$ & $  \ell_p / \ell $ & $  5 \times 10^{-3}$ \\
$\Pb$ & $  L (S_l - S_u)/(2S_l c \tau)\left\{1 - \beta_T (T_l - T_u)/\left[\beta_S (S_l - S_u)\right]\right\}$   & $  0.2$ \\
$\Pt$ & $ \left[T_l - T_u + \Gamma( S_l - S_u)\right]/(2 \tau)$ & $  1.2$ \\
$\kappa$ & $ (S_l + S_u)/(2S_l) -  \beta_T L / (c \beta_S S_l)$ &  $  1.2$ \\
\hline
$k_2$ & $\epstwo/\epsone$ & 1.7\\
$k_3$ & $\epsthree / \epsone$ & 1.4\\
$k_4$ & $\epsfour/ \epsone$ & 0.3\\
\end{tabular}
\end{center}
\end{table}
 
There are eight further dimensionless parameters in the system~\eqref{E:NonDim:mass}--\eqref{E:NonDim:thermal}, which are defined in table~\ref{T:Dimensionless_Parameters}. We adopt the naming convention that small parameters are denoted by $\epsilon_i$, with the exception of the dimensionless pycnocline thickness $\delta = \ell_p / \ell$, since this parameter is unique in defining a region of the solution, as we shall see. The other dimensionless variables, $\Pb$, $\Pt$, and $\kappa$, are treated as $\order{1}$ (table~\ref{T:Dimensionless_Parameters}).

As we shall see in \S\ref{S:Asymptotics}, $\Pb$ captures the dominant contribution to buoyancy deficit changes across the pycnocline. Its typical value is positive (table~\ref{T:Dimensionless_Parameters}), suggesting that buoyancy changes across the pycnocline are predominantly the result of variations in ambient salinity rather than variations in ambient temperature.

The final \blue{two} terms on the right-hand side of~\eqref{E:NonDim:buoyancy} \blue{(those in the square brackets)} describe\rout{s} the contribution of \rout{gradients }ambient temperature and salinity to variations in \blue{the} buoyancy deficit \blue{between the plume and the ambient}. Within this term, the parameters $\kappa$ and $\epsfour$ \blue{parametrise} \rout{describe }the \blue{size of the} depth-mean and depth-dependent contributions, respectively; the fact that $\epsfour \ll \kappa$ indicates that the depth-mean contribution plays the dominant role.

The parameter $\Pt$ describes the strength of changes in thermal driving across the pycnocline. Although we include salinity variations in the definition of $\Pt$, we note that $\Gamma (S_l - S_u)/(T_l - T_u) \ll 1$ (table~\ref{T:Constants}) and thus the parameter $\Pt$ is dominated by temperature variations in the ambient.

To make progress using asymptotic methods, we must specify the relative sizes of the small parameters $\lt$ and $\epsilon_i, ~i = 1\dots4$. Based on the typical values listed in table~\ref{T:Dimensionless_Parameters}, we assume that they all have the same asymptotic order, i.e. $\epstwo, \epsthree, \epsfour, \lt = \order{\epsone}$. To simplify the notation to a single small parameter ($\epsone$), we therefore introduce the $\order{1}$ parameters $k_i =\epsilon_i/ \epsone,~i = 2,3,4$ (values of the $k_i$ are also given in table~\ref{T:Dimensionless_Parameters}).

The problem~\eqref{E:NonDim:mass}--\eqref{E:NonDim:thermal} is closed by specifying a single boundary condition at the grounding line. Our assumption of negligible subglacial discharge is encoded by taking
\begin{equation}\label{E:NonDim:IC1}
\hat{U} =0,~\hat{D} = 0,
\end{equation}
at $\hat{X} = 0$. The boundary conditions on the buoyancy deficit and thermal driving are obtained by evaluating (the dimensionless version of) the similarity solution~\eqref{E:Similarity:SimilaritySol1}--\eqref{E:Similarity:SimilaritySol2} in the limit $\hat{X} \to 0$. To leading order in $\epsone$ this is
\begin{equation}\label{E:NonDim:IC2}
  \Delta\hat{\rho}(0) = \kappa,~\Delta \hat{T}(0) = 1  
\end{equation}
at $\hat{X} = 0$. Henceforth, hats are dropped, and all quantities are assumed dimensionless, unless otherwise stated.

\subsection{Example Solutions}\label{S:ExampleSolutions}

\begin{figure}
\centering
\includegraphics[width = \textwidth]{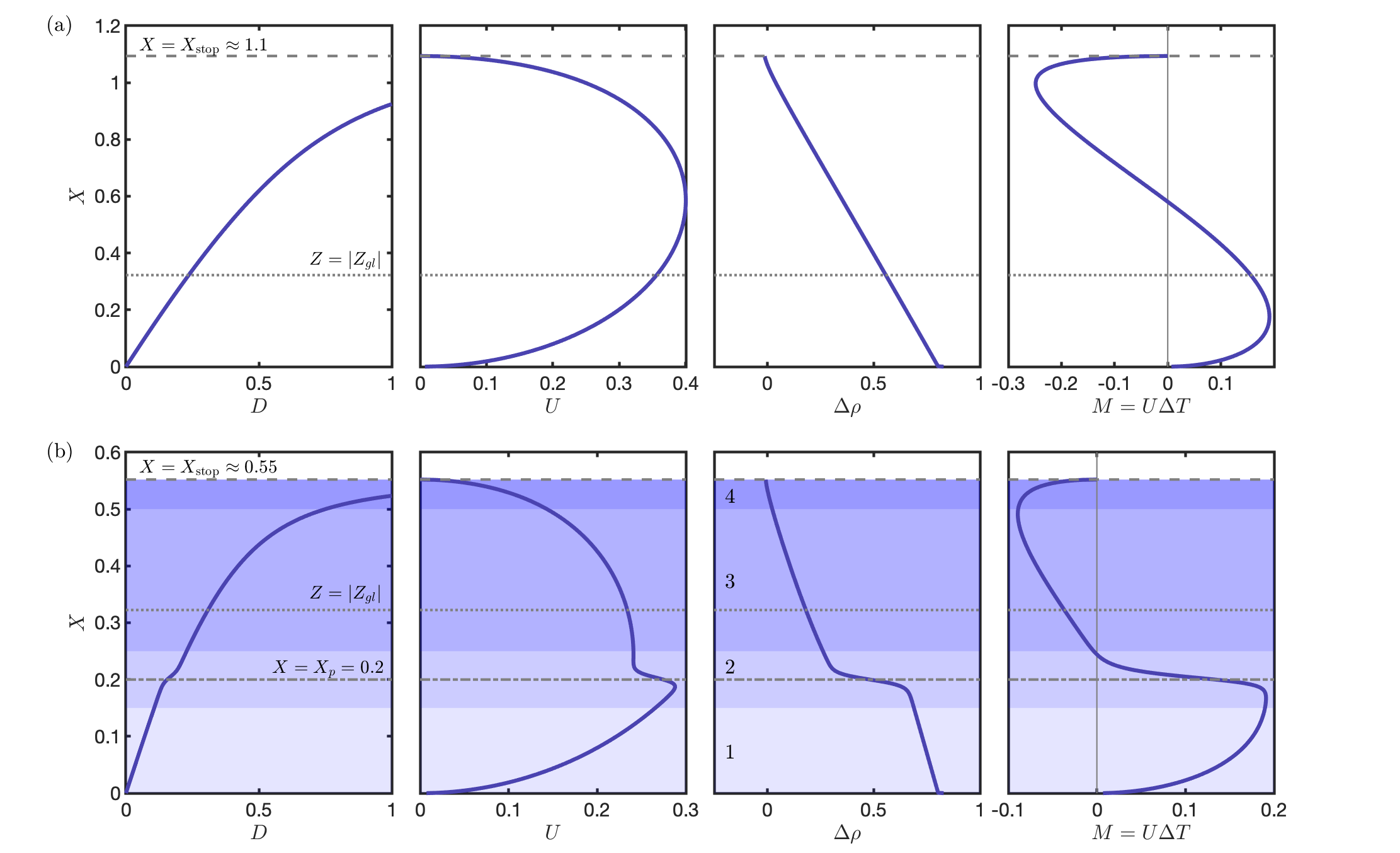}
\caption{Numerical solutions of dimensionless plume equations~\eqref{E:NonDim:mass}--\eqref{E:NonDim:thermal}. Solutions are shown for (a) no stratification, (all parameters as in table~\ref{T:Dimensionless_Parameters} except $\Pb = 0$, $\Pt = 0$, $\epsfour = 0$) and (b)  a typical two-layer ambient stratification (all parameters as in table~\ref{T:Dimensionless_Parameters}). \blue{In both cases, the panels indicate, from left to right: the plume thickness $D$, velocity $U$, buoyancy deficit $\Delta \rho$, and melt rate $M = U \Delta T$, where $\Delta T$ is the thermal driving. Solutions are plotted as a function of the along-shelf co-ordinate $X$, which is equal to the depth for the linear ice shelf base $Z_b(X) = X$ considered here.} \rout{Here we take a linear ice shelf base, $Z_b(X) = X$. }\blue{In both cases, t}\rout{T}he dashed horizontal line \blue{at the top of the figure} indicates $X_{\text{stop}}$, where the plume velocity reaches zero \blue{[note the different scales of the ordinate axis in (a) and (b)].} The \rout{dot-dashed} \blue{dotted} horizontal line indicates $Z = |Z_{gl}|$, the dimensionless depth of the grounding line, which restricts the solution domain in practice (see main text). In (b), the \rout{solid }\blue{dot-dashed} grey line indicates $X = X_p$, the centre of the pycnocline. Shading (and numbers in the third panel) indicate schematically the regions described in \S2\ref{S:ExampleSolutions} and introduced formally in \S\ref{S:Asymptotics}. }\label{fig:ExampleSols}
\end{figure}

Before proceeding with an asymptotic analysis of the model equations~\eqref{E:NonDim:mass}--\eqref{E:NonDim:thermal} it is instructive to consider typical solutions, which provide insight into the characteristic behaviour and the influence of ambient stratification on it. These solutions (and those presented later) are obtained numerically; to do so, we express~\eqref{E:NonDim:mass}--\eqref{E:NonDim:thermal} as a single matrix ODE, which is solved numerically in~\textsc{MATLAB} using the \texttt{ODE15s} routine. Software code to reproduce the calculations and figures in the paper is available at reference~\cite{PycnoclineCode}.

We consider first the limiting case in which there is no ambient stratification, taking $\Pb = 0, \Pt = 0, \epsfour = 0$ in~\eqref{E:NonDim:mass}--\eqref{E:NonDim:thermal}. Solutions in this case are shown in figure~\ref{fig:ExampleSols}a; the behaviour has been described in detail elsewhere~\cite[see][for example]{Jenkins1991JGeophysResOceans, Jenkins2011JPhysOcean, Magorrian2016JGeoResOcean, Hewitt2020AnnRevFlu}: briefly, the plume accelerates as it moves away from the grounding line, where it is strongly buoyant; the high velocities and strong thermal driving at depth result in high melt rates near the grounding line (figure~\ref{fig:ExampleSols}a, fourth panel). As the plume rises, it entrains salty, dense ambient water, leading to a retardation of the velocity (second panel), and the local freezing point increases; eventually (at $X \approx 0.5$ in figure~\ref{fig:ExampleSols}a) the thermal driving, and thus melt rate, becomes negative, indicating refreezing (fourth panel). Beyond this point, meltwater is being removed from the plume, further increasing its density (reducing its density deficit, second panel), in addition to continued entrainment of salty ambient water. The plume becomes negatively buoyant further downstream (at $X \approx 0.9$ in figure~\ref{fig:ExampleSols}a). It continues to move upwards, driven by its momentum but increasingly restrained by gravity; its velocity ultimately reaches zero, and the plume thickness diverges, at a point that we denote $X = X_{\text{stop}}$. Note that, although we show results for all $X < 1$ here, $X$ is restricted in practice by the grounding line depth: the subglacial plume can only extend as far as the sea level. For typical grounding line depths and constant ambient salinity, it is uncommon to encounter the region of refreezing (figure~\ref{fig:ExampleSols}a). 

With two-layer ambient stratification imposed (figure~\ref{fig:ExampleSols}b), the behaviour is identical below the pycnocline: the plume does not know about the pycnocline until it is very close to it. Across the pycnocline, however, each of the dependent variables experience a sharp change: as the plume enters the pycnocline it suddenly `sees' ambient conditions that are significantly colder and fresher. The plume leaves the region around the pycnocline with lower speed and thermal driving than in the unstratified ambient case, and both the point at which the plume reaches neutral buoyancy and $X_\text{stop}$ are reduced (for the values used in figure~\ref{fig:ExampleSols}, $X_{\text{stop}}$ in  the stratified ambient case is approximately half of its value in the unstratified ambient case).

%describe layers
The behaviour observed in the stratified ambient case can be divided qualitatively into four distinct regions, which are indicated schematically in figure~\ref{fig:ExampleSols}b. The first is adjacent to the grounding line, where the plume behaves as in the unstratified ambient case; the second region surrounds the pycnocline; the third lies above the pycnocline, where the plume decelerates and the buoyancy deficit reduces steadily; the fourth lies adjacent to $X_{\text{stop}}$, where the plume velocity transitions sharply to zero and the plume thickness diverges.  These four regions are defined formally in the asymptotic analysis, which we turn to now. 

\section{Asymptotic Analysis}\label{S:Asymptotics}
In this section, we present an asymptotic analysis of the model equations~\eqref{E:NonDim:mass}--\eqref{E:NonDim:thermal}, with a view to constructing an analytic approximation to the scaled melt rate $M = U \Delta T$. \blue{Readers who are primarily interested in the specification and performance of the approximation, rather than technical details of the asymptotic analysis, may wish to skip this section and proceed to \S\ref{S:MeltRate}.} Here we consider only the leading order behaviour: all the dependent variables discussed in this section can be considered to be leading order terms in an asymptotic expansion~\cite{HinchPerturbationMethods} in powers of $\epsone$. 

As part of this analysis, the four regions introduced descriptively in the previous section emerge naturally; in turn, we formally define each region, then make a variable scaling (if appropriate) and describe the structure of solutions to the resulting leading order equations, obtained by retaining only leading order terms in~\eqref{E:NonDim:mass}--\eqref{E:NonDim:thermal}. In each region, this analysis reveals where the solution to the leading order equations breaks down; by analysing the behaviour close to this point, we obtain matching conditions for the next region downstream. %As we shall see, each region has a different balance in the momentum equation to the region located upstream of it; these balances can therefore be used to characterize the regions and are listed in table~\ref{T:Balances}. 

Note that throughout this section, we use primes to denote derivatives with respect to $X$, and the argument of variables is $X$ unless specified.

%\begin{table}[h!]
%\caption{Nomenclature and momentum balances in the regions described in the asymptotic analysis of \S\ref{S:Asymptotics}.}
%\label{T:Balances}
%\begin{center}
%\begin{tabular}{lll}
%  Region  & Descriptive name & Leading order momentum balance    \\ \hline
%   1  & Below the pycnocline & Buoyancy, drag \\
%   2  & Pycnocline & Inertia, buoyancy, drag \\
%   3  & Above the pycnocline & Buoyancy, drag \\
%   4  & Stopping & Inertia,  buoyancy, drag
%\end{tabular}
%\end{center}
%\vspace{-4pt}
%\end{table}

\subsection{Region one: below the pycnocline}\label{S:Asymptotics:Region1}
The first region is immediately upstream of the grounding line,  with $X  <  X_p$,  $X= \order{1}$, and $X - X_p = \mathcal{O}(1)$. We refer to this as region one or, `below the pycnocline'. Here, the leading order equations are:
\begin{align}
%\dd{(DU)}{X} &= U \dd{Z_b}{X},\label{E:Region1:mass}\\
%0 &= D \Delta \rho \dd{Z_b}{X} - U^2, \label{E:Region1:mom}\\
%\dd{(DU\Delta \rho)}{X}  &=\lambda U \Delta T\label{E:Region1:buoyancy}\\
%0&= (1  - Z_b)\dd{Z_b}{X}U- U\Delta T - DU \dd{Z_b}{X}.\label{E:Region1:thermal}
(DU)' &= U Z_b',\label{E:Region1:mass}\\
0 &= D \Delta \rho Z_b' - U^2, \label{E:Region1:mom}\\
(DU\Delta \rho)'  &=\kappa U \Delta T\label{E:Region1:buoyancy}\\
0&= (1  - Z_b)Z_b' U- U\Delta T - DU Z_b'.\label{E:Region1:thermal}
\end{align}
\rout{Equation }\blue{The single term on the right-hand side of}~\eqref{E:Region1:mass} \blue{corresponds to entrainment, indicating that mass contributions from melting do not appear at leading order in region one. This confirms the assertion made in \S2\ref{S:Model:SimilaritySolution} that mass contributions from melting are unimportant in comparison with entrainment in proximity to the grounding line.}  \rout{demonstrates that contributions to mass conservation from melting are not important near the grounding line, as was suggested earlier in \S\ref{S:Model}. }

The system~\eqref{E:Region1:mass}--\eqref{E:Region1:thermal} is identical to that considered by~\cite{Lazeroms2019JPhysOcean}, albeit with a general ice shelf basal geometry shape; following~\cite{Lazeroms2019JPhysOcean}, we make progress by considering the flux $Q= DU$. Mass conservation~\eqref{E:Region1:mass} gives a first relationship between the flux and the velocity: $U = Q'/Z_b'$. A second relationship is obtained by eliminating $\Delta \rho$ from~\eqref{E:Region1:mom}--\eqref{E:Region1:buoyancy} and combining the result with~\eqref{E:Region1:thermal}: $[U^3/Z_b' - \kappa (1 - Z_b)Q]' = 0$. Eliminating the speed $U$ from these two relationships between the flux and velocity gives a second order ODE for $Q$; the solution to this ODE which satisfies the boundary condition~\eqref{E:NonDim:IC1} is 
\begin{equation}\label{E:Region1:flux_solution_integral}
Q(X) =  \left(\frac{2}{3}\right)^{3/2} \kappa^{1/2}I(X)^{3/2}, \quad \text{where} \quad I(X) =  \int_0^X \left[Z_b'(\xi)\right]^{4/3}\left[1 - Z_b(\xi)\right]^{1/3}~\mathrm{d}\xi.
\end{equation}
The plume velocity can then be determined using $U = Q'/Z_b'$ and~\eqref{E:Region1:flux_solution_integral},
\begin{equation}\label{E:Region1:velocity_solution}
U(X) =\left(\frac{2\kappa}{3}\right)^{1/2}\left[Z_b'(X)\right]^{1/3}\left[1 - Z_b(X)\right]^{1/3}I(X)^{1/2}.
\end{equation}
Finally, the plume thickness, buoyancy deficit, and thermal driving can be determined algebraically from~\eqref{E:Region1:mom} and~\eqref{E:Region1:thermal}. The particular form of these solutions is unimportant, and its statement is thus relegated to the supplementary information. The leading order contribution to the melt rate, which results from these, is stated explicitly in \S\ref{S:MeltRate}.

\blue{We make several remarks on these results: firstly,} \rout{Note that }the presence of the integral in~\eqref{E:Region1:flux_solution_integral} indicates that the flux, and thus melt rate, at an arbitrary point below the pycnocline depends on the geometry of the ice shelf base everywhere upstream of it. \blue{Secondly, we note that }\rout{I}\blue{i}n general, the integral in~\eqref{E:Region1:flux_solution_integral} cannot be evaluated analytically, but in the case special case of a planar geometry [$Z_b(X) = X$] \blue{it can be evaluated explicitly, in which case the solution reduces to the result of}~\cite{Lazeroms2019JPhysOcean}. \blue{Thirdly, }\rout{Note that }by taking the limit $X \to 0$ in this solution, we recover 
\begin{equation}\label{E:Region1:dimensionless_similarity}
D = \frac{2}{3}X, \quad U = \left(\frac{2\kappa}{3}\right)^{1/2}X^{1/2}, \quad \Delta \rho = \kappa, \quad \Delta T = 1,
\end{equation} 
which are the leading order terms of an asymptotic expansion in $\epsone$ of the similarity solution~\eqref{E:Similarity:SimilaritySol1}--\eqref{E:Similarity:SimilaritySol2} rescaled according to~\eqref{E:NonDim:NonDimVar1}--\eqref{E:NonDim:NonDimVar2}, as expected. \blue{Finally, we }\rout{N}\blue{n}ote that in the absence of ambient stratification, the solution applies for all $X> 0$ where $1 - Z_b(X) = \mathcal{O}(1)$ [the solution~\eqref{E:Region1:velocity_solution} breaks down as $Z_b(X) \to 1$, where $U \to 0$ but this point is well beyond typical depths, as discussed in \S2\ref{S:ExampleSolutions}]. We shall return to this point in \S\ref{S:Numerics}.

\subsection{Region two: pycnocline}\label{S:Asymptotics:Region2}
The analysis of the previous section remains valid while $X - X_p = \mathcal{O}(1)$. When $X - X_p = \order{\delta}$, this analysis breaks down; to analyze the behaviour in this region, we set $X = X_p + \lt \zeta$, where $\zeta = \mathcal{O}(1)$ in the model equations~\eqref{E:NonDim:mass}--\eqref{E:NonDim:thermal}. The resulting system, referred to as the `pycnocline-rescaled plume equations' is stated in the supplementary. Retaining only leading order terms in the rescaled pycnocline-rescaled plume equations, we obtain
\begin{align}
\dd{(DU)}{\zeta} &=0,		\label{E:PycnoclineRegion:Mass}	\\
\cone \dd{(DU^2)}{\zeta} &=  D\Delta \rho Z_b'(X_p) - U^2,	\label{E:PycnoclineRegion:Mom}	\\
\dd{(DU\Delta \rho)}{\zeta} &= -\Pb \mathrm{sech}^2(\zeta)Z_b'(X_p)DU  \label{E:PycnoclineRegion:Buoyancy}		\\
\ctwo \dd{(DU\Delta T)}{\zeta} &= \left\{1 - Z_b(X_p) - \Pt\left[1 + \tanh(\zeta)\right]  -D\right\}UZ_b'(X_p) - U\Delta T.\label{E:PycnoclineRegion:Themal}
\end{align}
where the constants $c_i = \epsilon_i / \delta = \order{1},~i = 1, 2$ have been introduced for clarity. The system~\eqref{E:PycnoclineRegion:Mass}--\eqref{E:PycnoclineRegion:Themal} is to be solved on $-\infty < \zeta < \infty$. [For the numerical solution presented below, we solve~\eqref{E:PycnoclineRegion:Mass}--\eqref{E:PycnoclineRegion:Themal} on $-\zeta_0 < \zeta < \zeta_0$ where $|\zeta_0| \ll \epsone^{-1}$.]

Matching with region one simply involves evaluating the region one solution at $X = X_p$. These quantities are denoted by a subscript `in'\rout{i} \blue{(the `in' referring to this being the solution upon entry into the pycnocline region)}, i.e. we require
\abcdeqn{E:PycnoclineRegion:FarField}{
U \to U_{\text{in}}, \quad D \to D_{\text{in}}, \quad \Delta \rho \to \Delta \rho_{\text{in}}, \quad \Delta T \to \Delta T_{\text{in}} \quad \text{as}~\zeta \to -\infty.}
We state $U_{\text{in}}$, $D_{\text{in}}$, $\Delta \rho_{\text{in}}$, and $\Delta T_{\text{in}}$ explicitly in the supplementary.

To make progress, we first note that, from~\eqref{E:PycnoclineRegion:Mass}, flux is conserved across the pycnocline:
\begin{equation}\label{E:PycnoclineRegion:Q_conserved}
Q(\zeta) = Q_{\in} = D_{\in} U_{\in}  =\left(\frac{2}{3}\right)^{3/2} \kappa^{1/2}I(X_p)^{3/2}.
\end{equation}
Inserting~\eqref{E:PycnoclineRegion:Q_conserved} into~\eqref{E:PycnoclineRegion:Buoyancy} gives in an ODE for $\Delta \rho$, which can be solved alongside~\eqref{E:PycnoclineRegion:FarField}c to give
\begin{equation}\label{E:PycnoclineRegion:Deltarho_solution}
\Delta \rho = \Delta \rho_{\in} - \Pb Z_b'(X_p) \left[1 + \tanh(\zeta)\right].
\end{equation} 

Note that on the $\order{1}$ length scale of the whole ice shelf, the presence of a pycnocline essentially introduces a sharp change in each of the dependent variables. Since we know the value of dependent variables as the plume approaches the pycnocline, the \textit{change} in dependent variables across the pycnocline region is the key information we must glean from the analysis in this section. With this in mind, we note that the solution~\eqref{E:PycnoclineRegion:Deltarho_solution} gives $\Delta \rho \to \Delta \rho_{\in} - 2 \Pb Z_b'(X_p)  \eqcolon \Delta \rho_\out$ as $\zeta \to \infty$, i.e. the buoyancy deficit experiences a change
\begin{equation}\label{E:PycnoclineRegion:Deltarho_change}
\left[\Delta \rho\right]_{\text{pyc}} = \Delta \rho_\out - \Delta \rho_\in = - 2 \Pb Z_b'(X_p) < 0
\end{equation}
across the pycnocline region. Expression~\eqref{E:PycnoclineRegion:Deltarho_change} elucidates our description of $\Pb$ in \S\ref{S:Model} as a key parametric control on changes in buoyancy across the pycnocline.

\begin{figure}
\centering
\includegraphics[width = \textwidth]{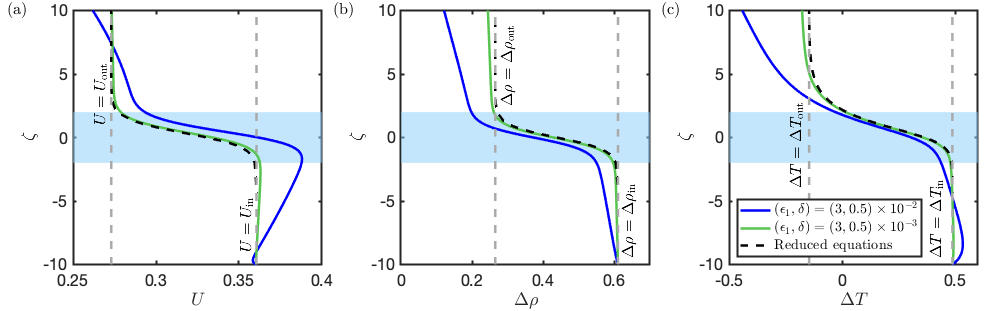}
\caption{Comparison between numerical solutions of the \blue{pycnocline-rescaled plume model} equations \rout{rescaled to the pycnocline}\blue{(solid blue and green curves)} and their \rout{leading order reduction }\blue{reduced form}~\eqref{E:PycnoclineRegion:Mass}--\eqref{E:PycnoclineRegion:Themal} \blue{(black dashed curves)}. \blue{For each,} \rout{W}\blue{w}e show (a) \blue{the} plume speed $U$, (b) \blue{the} buoyancy deficit $\Delta \rho$, and (c) \blue{the} thermal driving $\Delta T$ \blue{as a function of $\zeta = (X - X_p)/\delta$}. \blue{Blue and green solid}\rout{ Purple} curves correspond to the \blue{different values of the asymptotic parameters $\epsone$ and $\epstwo$, as follows [and indicated by the legend in (c)]:} \rout{full equations with} \blue{blue:} $\epsone = 3\times10^{-2}, ~\lt = 5\times10^{-3}$, i.e. as in table~\ref{T:Dimensionless_Parameters}, and \blue{green:} $\epsone = 3\times10^{-3},~ \lt = 5\times10^{-4}$, i.e. both asymptotic parameters reduced by a factor of ten\blue{, compared to the blue curves,} thus preserving their ratio. \rout{Black dashed curves correspond to the solution of the leading order equations.} Also plotted, as grey dashed lines, are the quantities $f_{\in}$ and $f_{\out}$, for $f = U,~ \Delta \rho, ~\Delta T$ as appropriate (defined in the main text). Solutions here use a constant ice shelf basal slope, $Z_b(X) = X$ and $c_1, c_2, k_3,k_4, \kappa$ have the values listed in table~\ref{T:Dimensionless_Parameters}. The coloured boxes indicate the region $-2 \leq \zeta  \leq 2$, across which the majority of the variation in the solution around $\zeta = 0 $ occurs.}\label{fig:PycnoclineAsymptotics}
\end{figure}

For the velocity $U$, we first combine~\eqref{E:PycnoclineRegion:Mom},~\eqref{E:PycnoclineRegion:Q_conserved}, and~\eqref{E:PycnoclineRegion:Deltarho_solution} to obtain the following ODE: 
\begin{equation}\label{E:PycnoclineRegion:U_ODE}
\cone Q_\in U \dd{U}{\zeta} = Q_\in Z_b'(X_p)\left\{\Delta \rho_\in - \Pb Z_b'(X_p) \left[1 + \tanh(\zeta)\right]\right\} - U^3.
\end{equation}
Unfortunately,~\eqref{E:PycnoclineRegion:U_ODE} cannot be solved analytically. However, for $\zeta \gg 1$ it can be approximated by
\begin{equation}\label{E:PycnoclineRegion:U_ODE_farfield}
\cone Q_\in U \dd{U}{\zeta} = Q_\in Z_b'(X_p)\Delta \rho_\out - U^3,
\end{equation}
which has a single fixed point 
\begin{equation}\label{E:PycnoclineRegion:U_Limit}
U = \left[ Q_\in Z_b'(X_p)\Delta \rho_\out\right]^{1/3}  \eqcolon U_\out.
\end{equation}
By noting that $\mathrm{d}U/\mathrm{d}\zeta > 0$  for $U < U_\out$ and $\mathrm{d}U/\mathrm{d}\zeta < 0$  for $U > U_\out$ in~\eqref{E:PycnoclineRegion:U_ODE_farfield}, we deduce that this fixed point is globally stable; the solution to~\eqref{E:PycnoclineRegion:U_ODE} therefore has $U \to U_\out$ as $\zeta \to \infty$ (figure~\ref{fig:PycnoclineAsymptotics}b). The change in plume velocity across the pycnocline region is
\begin{equation}\label{E:PycnoclineRegion:U_change}
\left[U\right]_{\text{pyc}} = U_\out - U_\in = \left[Q_\in Z_b'(X_p)\right]^{1/3} \left(\Delta \rho_\out^{1/3} - \Delta \rho_\in^{1/3}\right) < 0,
\end{equation}
where the sign of the inequality comes from~\eqref{E:PycnoclineRegion:Deltarho_change}. Mass conservation~\eqref{E:PycnoclineRegion:Q_conserved} gives the corresponding change in plume thickness,
\begin{equation}\label{E:PycnoclineRegion:D_change}
\left[D\right]_{\text{pyc}} = Q_\in \left(\frac{1}{U_\out} - \frac{1}{U_\in}\right) > 0.
\end{equation}

%take-aways for thermal driving
The thermal driving behaves similarly to the velocity: inserting~\eqref{E:PycnoclineRegion:Q_conserved} into~\eqref{E:PycnoclineRegion:Themal} gives the ODE [which is coupled to~\eqref{E:PycnoclineRegion:U_ODE}]
\begin{equation}\label{E:PycnoclineRegion:DeltaT_ODE}
\ctwo Q_\in \dd{\Delta T}{\zeta} = \left\{1 - Z_b(X_p) - \Pt\left[1 + \tanh(\zeta)\right]\right\}UZ_b'(X_p) - U\Delta T - Q_\in Z_b'(X_p), 
\end{equation}
which must also be solved numerically. For $\zeta \gg 1$,~\eqref{E:PycnoclineRegion:DeltaT_ODE} has a single fixed point,
\begin{equation}\label{E:PycnoclineRegion:DeltaT_Limit}
\Delta T = \frac{1}{U_\out}\left\{\left[1 - Z_b(X_p) - 2\Pt\right]U_\out Z_b'(X_p) - Q_\in Z_b'(X_p)\right\} \eqcolon \Delta T_\out.
\end{equation}
This fixed point is also globally stable, and thus $\Delta T \to \Delta T_\out$ as $\zeta \to \infty$ (figure~\ref{fig:PycnoclineAsymptotics}c); the thermal driving changes across the pycnocline region by an amount 
\begin{equation}\label{E:PycnoclineRegion:DeltaT_change}
\left[T\right]_{\text{pyc}} = \Delta T_\out - \Delta T_\in < 0.
\end{equation}

The sign of the `jump conditions'~\eqref{E:PycnoclineRegion:Deltarho_change},~\eqref{E:PycnoclineRegion:U_change},~\eqref{E:PycnoclineRegion:D_change}, and~\eqref{E:PycnoclineRegion:DeltaT_change} are in qualitative agreement with the example solution presented in \S\ref{S:Model}. To go beyond this qualitative comparison, in figure~\ref{fig:PycnoclineAsymptotics} we present a comparison between the pycnocline-rescaled plume equations and solutions of the corresponding leading order equations~\eqref{E:PycnoclineRegion:Mass}--\eqref{E:PycnoclineRegion:Themal}  [for $U$ and $\Delta T$, this is the numerical solution of~\eqref{E:PycnoclineRegion:U_ODE} and~\eqref{E:PycnoclineRegion:DeltaT_ODE}, respectively, and for $\Delta \rho$, this is the solution~\eqref{E:PycnoclineRegion:Deltarho_solution}]. Solutions of the leading order equations show good agreement with the full equations, albeit with a drift at larger $\zeta$ that is primarily the result of variations on the outer variable length scale, and a discrepancy at large negative $\zeta$ that results from applying  initial conditions at $\zeta = -\zeta_0$ rather than at $\zeta = -\infty$. Most importantly for the approximation to the melt rate that is presented in \S\ref{S:MeltRate}, the jump conditions~\eqref{E:PycnoclineRegion:U_change},~\eqref{E:PycnoclineRegion:Deltarho_change}, and~\eqref{E:PycnoclineRegion:DeltaT_change} do an excellent job at predicting the change in the dependent variables in the region around $\zeta = 0$.

Before moving on to analyze the  region above the pycnocline, we note that the results of this section are only valid provided that $\Delta \rho$ remains $\order{1}$ throughout. In particular, this means that if $\Delta \rho_{\in} < 2 \Pb Z_b'(X_p)$ [see equation~\eqref{E:PycnoclineRegion:Deltarho_solution}] the analysis will break down and a sub-layer within the pycnocline is needed to proceed. This analysis is similar to that presented in \S\ref{S:Asymptotics}(d) so we do not discuss it further here, other than to note that it would indicate that the plume speed reaches zero within the pycnocline, and the plume therefore detaches from the ice shelf base. This scenario may be expected to occur if the salinity stratification is particularly strong (high $\Pb$) or if the pycnocline is high in the water column (the buoyancy deficit has reduced significantly in region one, resulting in low $\Delta \rho_{\in}$). Using typical parameter values for Antarctica, we do not expect separation within the pycnocline to be predicted by the plume model considered here [see \S\ref{S:Model}(d)], but note that observation in this region commonly point towards the presence of meltwater within the pycnocline. This is consistent with the picture presented earlier of progressive detrainment from the plume into the stratified ambient, which is not included in our model. For ice shelves in Greenland, however, where shelf bases are typically much steeper than in Antarctica and subglacial discharge may be significant, plumes have been observed to intrude into the ambient layer within the pycnocline~\cite{Straneo2011NatureGeo}. We speculate that the criterion $\Delta \rho_\out < 0$ may provide a reasonable first guess of whether a plume will separate from the ice shelf base within the pycnocline or not.

\subsection{Region three: above the pycnocline}\label{S:Asymptotics:Region3}
We now turn to the region above the pycnocline,  where $X > X_p$ and $X = \order{1}$. We refer to this as region three. The leading order equations in region three read
%\begin{align}
% (DU)' &= U Z_b',\label{E:Region3:mass}\\
%0 &= D \Delta \rho Z_b' - U^2, \label{E:Region3:mom}\\
%(DU\Delta \rho)'  &=\kappa U \Delta T\label{E:Region3:buoyancy}\\
%0&= (1  - 2\Pt -  Z_b)Z_b'U- U\Delta T - DU Z_b'.\label{E:Region3:thermal}
%\end{align}
\begin{align}
 (DU)' &= U Z_b', &
0 &= D \Delta \rho Z_b' - U^2, \label{E:Region3:mass}\\
(DU\Delta \rho)'  &=\kappa U \Delta T, &
0&= (1  - 2\Pt -  Z_b)Z_b'U- U\Delta T - DU Z_b'.\label{E:Region3:thermal}
\end{align}

The system~\eqref{E:Region3:mass}--\eqref{E:Region3:thermal} is almost identical to the leading order equations in region one [\eqref{E:Region1:mass}--\eqref{E:Region1:thermal}]: buoyancy balances drag in the momentum equation, mass flux is driven solely by entrainment, and buoyancy variations are controlled by melting only. However, conservation of thermal driving [second of~\eqref{E:Region3:thermal}] reflects a different ambient temperature to~\eqref{E:Region1:thermal}, which results from having crossed the pycnocline.

Equations~\eqref{E:Region3:mass}--\eqref{E:Region3:thermal} are to be solved for $X > X_p$, alongside boundary conditions obtained by matching to region two:
\begin{equation}\label{E:Region3:MatchingConditions}
U = U_\out, \quad D = \frac{Q_\in}{U_\out}, \quad \Delta \rho = \Delta \rho_\out, \quad \Delta T = \Delta T_\out \quad \text{at}~X = X_p.
\end{equation}

We proceed as in \S3\ref{S:Asymptotics:Region1}: from the first of~\eqref{E:Region3:mass}, the flux $Q = DU$ and velocity $U$ are again related by $U =Q'/Z_b'$,
which, when combined with the remaining three equations in~\eqref{E:Region3:mass}--\eqref{E:Region3:thermal}, gives a second order ODE for $Q$. Integrating this ODE once alongside~\eqref{E:Region3:MatchingConditions} gives
\begin{equation}\label{E:Region3:Q_ODE}
%\left(Q'\right)^3 =\kappa \left(Z_b'\right)^4 \left\{\left(1 - Z_b\right)Q - 2\Pt\left(Q - Q_\in\right) -\left[1 - Z_b(X_p)\right]Q_\in\right\} + \frac{U_\out^3}{Z_b'(X_p)}\left(Z_b'\right)^4.
\frac{\left[Q'(X)\right]^3}{\left[Z_b'(X)\right]^4} = \kappa \left\{ \left[1 - Z_b(X) - 2P_T\right] Q(X) - \left[1 - Z_b(X_p) - 2P_T\right]Q_\text{in}\right\} + \frac{U_\text{out}^3}{Z_b'(X_p)}
\end{equation}

%can't solve this numerically
Unlike the corresponding ODE for region one, equation~\eqref{E:Region3:Q_ODE} cannot be solved exactly, i.e. it is not possible to express the leading order contribution to the melt rate analytically in region three. Whilst this is problematic for constructing an approximation to the melt rate, we describe one possible workaround in \S\ref{S:MeltRate}, which exploits two pieces of information that we are able to glean from~\eqref{E:Region3:Q_ODE}: firstly, we note that from mass conservation~\eqref{E:Region3:mass}, the flux $Q$ must be increasing (the velocity $U$ is positive). $Q$ will therefore continue to increase until it reaches a critical point $X = X_c$, where $Q' \to 0$ and the leading order equations~\eqref{E:Region3:mass}--\eqref{E:Region3:thermal} break down. From~\eqref{E:Region3:Q_ODE}, we see that the flux at this point, $Q_c = Q(X_c)$, satisfies 
\begin{equation}
\left[1 - Z_b(X_c)\right]Q_c - 2\Pt\left(Q_c - Q_\in\right) -\left[1 - Z_b(X_p)\right]Q_\in +  \frac{U_\out^3}{\kappa Z_b'(X_p)} = 0.
\end{equation}

Secondly, it is instructive to analyse the behaviour as $X \to X_c$. As we show in the supplementary information, the following behaviour holds in the limit $X \to X_c^-$:
\begin{align}
U &\sim \kappa^{1/3} Z_b'(X_c)^{2/3} Q_c^{1/3}(X_c - X)^{1/3}, & 
D &\sim \kappa^{-1/3} Z_b'(X_c)^{-2/3} Q_c^{2/3}(X_c - X)^{-1/3}, \label{E:Region3:UD_asym} \\
\Delta \rho &\sim  \kappa Z_b'(X_c) (X_c - X), &
 \Delta T &\sim -\kappa^{-1/3} Z_b'(X_c)^{1/3} Q_c^{2/3}(X_c - X)^{-1/3}.\label{E:Region3:drhodT_asym}
 \end{align}
The asymptotic results~\eqref{E:Region3:UD_asym}--\eqref{E:Region3:drhodT_asym} also inform the matching conditions on the region around $X_c$, which we turn to now.

%Before moving on to analyse Region 5, we note that, as we shall see in \S\ref{S:Numerics}, the point $X_c$ is only located below sea level (where it is physically relevant) in ice shelves with uncommon parameters -- corresponding to very deep grounding lines or very cold ambient conditions, for example. We shall also see that the melt rate is small ($\order{\epsone}$) for $X \gtrsim X_c$: the behaviour above $X_c$ is not important for the leading order melt rate. However, we include the analysis for completeness, because of its relevance to other physical systems in which thin flows experience a competition between buoyancy and drag (such as Katabatic winds), and because the results of the analysis have implications for the 'layered-plume' hypothesis proposed by~\cite{Magorrian2016JGeoResOcean}. 

\subsection{Region four: stopping region}\label{S:Asymptotics:Region4}
%scaled variables for this region
\newcommand{\U}{\mathcal{U}}
\newcommand{\D}{\mathcal{D}}
\newcommand{\p}{\Delta \varrho}
\renewcommand{\t}{\Delta \mathcal{T}}

The analysis of the previous section breaks down as $X \to X_c$, where $U \to 0$. To analyze the behaviour around $X = X_c$, referred to as the `stopping region', we introduce the rescaled spatial variable $\chi$. The appropriate spatial scaling for a dominant balance in~\eqref{E:NonDim:mass}--\eqref{E:NonDim:thermal} is achieved by setting 
\begin{equation}\label{E:Region4:X_scaling}
X = X_c + \epsone^{3/4}\chi,
\end{equation}
where $\chi = \order{1}$. The behaviour~\eqref{E:Region3:UD_asym}--\eqref{E:Region3:drhodT_asym} then informs us of the size of the dependent variables; we introduce appropriately scaled variables
\begin{equation}\label{E:Region4:var_scaling}
U = \epsone^{1/4}\U, \quad D = \epsone^{-1/4}\D, \quad \Delta \rho = \epsone^{3/4}\p, \quad  \Delta T = \epsone^{-1/4} \t.
\end{equation}

Inserting~\eqref{E:Region4:X_scaling}--\eqref{E:Region4:var_scaling} into the model equations~\eqref{E:NonDim:mass}--\eqref{E:NonDim:thermal}, and retaining only leading order terms gives
\begin{align}
\dd{(\D\U)}{\chi} &=0, &
\dd{(\D \U^2)}{\chi} &=  \D \p Z_b'(X_c) -\U^2,\label{E:Region4:mass_scaled}\\
\dd{(\D\U\p)}{\chi} &=\kappa  \U\t, &
k_2 \dd{(\D \U \t)}{\chi} &=- \U \t - \D \U Z_b'(X_c).\label{E:Region4:thermal_scaled}
\end{align}

Matching with region three requires that the solution to~\eqref{E:Region4:mass_scaled}--\eqref{E:Region4:thermal_scaled} satisfies
\begin{align}
\U &\sim \kappa^{1/3}Z_b'(X_c)^{-2/3}Q_c^{1/3} (-\chi)^{1/3}, &  \D &\sim \kappa^{-1/3} Z_b'(X_c)^{-2/3} Q_c^{2/3}(-\chi)^{-1/3}\label{E:Region4:far_field1},\\
\p &\sim - \kappa Z_b'(X_c) \chi, & \t &\sim -\kappa^{-1/3} Z_b'(X_c)^{1/3} Q_c^{2/3}(-\chi)^{-1/3}\label{E:Region4:far_field2}
\end{align}
as $\chi \to -\infty$. 

\blue{A full analysis of this problem is beyond the scope of this paper. However, in the supplementary information, we show explicitly that the plume terminates within this region and that, in addition, the solution for the rescaled velocity $\U$ in~\eqref{E:Region4:mass_scaled}--\eqref{E:Region4:thermal_scaled} can be approximated by} 
\begin{equation}
     \U =  \kappa^{1/3}Z_b'(X_c)^{1/3}Z_b'(X_c)^{2/3} \left(\chi - \frac{C}{\kappa Z_b'(X_c)}\right)^{1/3},
\end{equation}
\blue{where $C$ is a constant that emerges from analysing the higher order contributions to $\Delta \rho$ as $X \to X_c$. However, we do not dwell on this behaviour because it is not directly used in constructing our approximation to the melt rate, which we turn to now.}

\section{Approximating the melt rate}\label{S:MeltRate}
In this section, we describe our analytic approximation $M_p(X)$ to the melt rate $M(X) = U(X)\Delta T(X)$ that emerges from the model equations~\eqref{E:NonDim:mass}--\eqref{E:NonDim:thermal}. This approximation builds upon the asymptotic analysis of \S\ref{S:Asymptotics}, and we consider each of the four regions identified in turn. 

\subsection{Region one}
Recall that in region one, the solution to the leading order equations can be expressed analytically; our approximation $M_p$ therefore takes values specified by the corresponding leading order contribution to $U\Delta T$:
\begin{equation}\label{E:MeltRate:region1}
M_{p} = M_{p,1}(X) = \left(\frac{2\kappa}{3}\right)^{1/2}Z_b'(X)\left[I(X)\right]^{1/2}\left\{\left[1 - Z_b(X)\right]^{1/3}\left[Z_b'(X)\right]^{1/3} - \frac{2}{3}I(X) \right\} %\coloneqq - Z_b'(X)\left\{\frac{2}{3}\int_0^X \kappa^{1/3}\left[Z_b'(\xi)\right]^{4/3}\left[1 - Z_b(\xi)\right]^{1/3}~\mathrm{d}\xi\right\}^{3/2} + \\
%\left(\frac{2\kappa}{3}\right)^{1/2}\left[Z_b'(X)\right]^{4/3}\left[1 - Z_b(X)\right]^{4/3} \left\{\int_0^X \left[Z_b'(\xi)\right]^{4/3}\left[1 - Z_b(\xi)\right]^{1/3}~\mathrm{d}\xi\right\}^{1/2},
\end{equation}
where $I(X)$ is given in equation~\eqref{E:Region1:flux_solution_integral}. The approximation~\eqref{E:MeltRate:region1} is valid for $0 < X < X_p - N_l \lt $, where $N_l = \order{1}$ is introduced to account for the finite extent of the pycnocline: we define the pycnocline region in  the approximation constructed here as $X_p - N_l \lt < X < X_p + N_l \lt$, i.e. the quantity $2 N_l$ describes the number of (dimensionless) pycnocline length scales required for the solution to transition between the constant values of the velocity and thermal driving on either side of the pycnocline.  We must make a choice for $N_l$; in what follows we take $N_l = 2$, a choice that is informed by the solutions of equations~\eqref{E:PycnoclineRegion:Mass}--\eqref{E:PycnoclineRegion:Themal}, shown in figure~\ref{fig:PycnoclineAsymptotics}. There it can be seen that the majority of the rapid change close to the centre of the pycnocline occurs over a length bounded by four dimensionless pycnocline length scales (the light blue boxes in figure~\ref{fig:PycnoclineAsymptotics} have height $4\lt$, when measured in terms of the outer variable $X$). \blue{The results presented here are insensitive to the value of $N_l$, provided that it is not too large: for $N_l \gtrapprox 4$, the pycnocline region takes up a disproportionately large portion of the water column, resulting in large errors not only in this region, but also in those regions above it, which rely on the solution in the pycnocline region.}

The integrals in~\eqref{E:MeltRate:region1} must be evaluated numerically in general, but accurate analytic approximations are readily computed provided that the ice shelf basal geometry is known. In the case that the ice shelf base has a constant slope, the approximation~\eqref{E:MeltRate:region1} reduces to the solution described by Lazeroms et al.~\cite{Lazeroms2019JPhysOcean}:
\begin{equation}\label{E:MeltRate:LazeromsMelt}
M_{\text{L19}}(X) = \frac{\kappa^{1/2}}{2\sqrt{2}}\left[1 - (1 - X)^{4/3}\right]^{1/2}\left[3(1-X)^{4/3} - 1\right].
\end{equation}
We henceforth refer to the dimensional form of~\eqref{E:MeltRate:LazeromsMelt} as the `L19 approximation'. 

Lazeroms et al.~\cite{Lazeroms2019JPhysOcean} also describe an ad-hoc method designed to account for non-constant basal slopes, in which all factors of the aspect ratio in the L19 approximation are replaced by the local slope of the ice shelf base, and is equivalent to multiplying~\eqref{E:MeltRate:LazeromsMelt} by a factor of $Z_b'(X)^{3/2}$; the corresponding approximation to the melt rate is
\begin{equation}\label{E:MeltRate:LazeromsMelt_adhoc}
M_{\text{L19AH}}(X) = \frac{\kappa^{1/2}}{2\sqrt{2}}\left[Z_b'(X)\right]^{3/2}\left[1 - (1 - X)^{4/3}\right]^{1/2}\left[3(1-X)^{4/3} - 1\right].
\end{equation}
The dimensional form of~\eqref{E:MeltRate:LazeromsMelt_adhoc} is referred to as the `L19AH approximation' henceforth. Lazeroms et al.~\cite{Lazeroms2019JPhysOcean} demonstrated that, and we show in \S\ref{S:Numerics}, that this simple method is reasonably effective at accounting for a non-constant basal slope.

The two-dimensional extensions of the L19~\eqref{E:MeltRate:LazeromsMelt} and L19AH~\eqref{E:MeltRate:LazeromsMelt_adhoc} approximations represent the current state of the art in plume-physics based melt rate parametrizations and therefore act as one benchmark against which we shall assess the approximation constructed in this section.

\subsection{Region two}
Unlike in region one, the leading order equations for region two~\eqref{E:PycnoclineRegion:Mass}--\eqref{E:PycnoclineRegion:Themal} do not have an analytic solution. We therefore construct our approximation based on expressions~\eqref{E:PycnoclineRegion:U_change} and~\eqref{E:PycnoclineRegion:DeltaT_ODE} for the change in $U$ and $\Delta T$ across a relatively slender pycnocline: we linearly interpolate according to these values, giving
\begin{equation}\label{E:MeltRate:regions2}
M_{p} = M_{p,2}(X) \coloneqq \left[U_{\out} + \left[U\right]_{\text{pyc}} \frac{X - (X_p + N_l \lt)}{2N_l \lt}\right]  \left[\Delta T_{\out} +  \left[\Delta T\right]_{\text{pyc}} \frac{X - (X_p + N_l \lt)}{2N_l \lt}\right] , 
\end{equation}
for $ X_p - N_l \lt  < X < X_p + N_l \lt$. [Recall that $U_\out$, $\left[U\right]_{\text{pyc}}$, $\Delta T_\out$, and $\left[\Delta T\right]_{\text{pyc}}$ are set out explicitly in equations~\eqref{E:PycnoclineRegion:U_Limit}, \eqref{E:PycnoclineRegion:U_change}, \eqref{E:PycnoclineRegion:DeltaT_Limit}, and~\eqref{E:PycnoclineRegion:DeltaT_change}, respectively.] Although the situation is not expected for parameter values appropriate for Antarctica [as discussed in \S3\ref{S:Asymptotics:Region2}], in the case that $\Delta \rho_\out < 0$, we apply~\eqref{E:MeltRate:regions2} only as far as $M_{p,2} = 0$ and take $M_{p} = 0$ downstream of this point.

\subsection{Region three}\label{S:MeltRate:R3}
Extra care must be taken for region three, owing to the dearth of information about the leading order solution in this region: our knowledge is limited to understanding that the velocity decreases until it reaches zero at an a priori unknown point $X_c$, and a description of the behaviour of solutions close to $X_c$ [in particular, we showed that $U \sim (X_c - X)^{1/3}$ as $X \to X_c$]. In this section, we describe a way to construct an appropriate approximation to the melt rate which exploits the information available, but stress that we make several ad hoc choices, and the construction presented below is by no means unique.

Our strategy for this region, $X_p + 2N_l \lt < X < X_c$, is to split it into two further regions: a lower part $X_p + 2 N_l < X \leq X^*$ and an upper part $X^* < X < X_c$. We choose to take $X^*$ to be \blue{smaller of $X_{\text{fr}}$, the value of $X$ that corresponds to the ice shelf front, and} the point at which the velocity has decayed to some fraction $0 < f < 1$ of $U_\out$, the velocity of the plume when it enters region three from below \rout{(assuming for the moment that such a point exists).}

For the lower part, $X_p  + 2N_l \lt < X \leq X^*$, we exploit the proximity to the pycnocline to artificially construct a small parameter, and seek an asymptotic expansion of the pertinent variables in this parameter. For the upper half, $X^* < X < X_c$, we mimic the behaviour as $X \to X_c$; in doing so we are able to simultaneously describe the behaviour in this region whilst also determining the values of $X_c$ and $Q_c = Q(X_c)$.

In more detail, for the lower half we introduce $\varepsilon = X^* - X_p$,
which we \rout{consider} \blue{shall assume} to be a small, positive parameter. \blue{We verify \textit{a posteriori}, once $X^*$ has been determined as outlined below, that $\varepsilon \ll 1$. Note, however, that because the vertical lengthscale chosen in \S2\ref{S:Model:NonDim} is typically much larger than the grounding line depth, we expect to have $X_{\text{fr}} < 1$ and thus $\varepsilon < 1$. } 

We then set $X = X_p + \varepsilon Y$, where  $Y = \order{1}$, in the ODE~\eqref{E:Region3:Q_ODE} for the flux $Q$ in region three, and attempt to account for variations in $Q$ away from $Q_{\text{in}} = Q(X_p)$ by expanding in powers of $\varepsilon$:
\begin{equation}\label{E:MeltRate:Q_expansionR3}
Q = Q_{3l} \coloneqq   Q_\in  + \sum_{i = 1}^{\infty} \varepsilon^i Q_i(Y), \qquad Q_i \sim \order{1}.
 \end{equation}
Equating powers of $\varepsilon$ leads to a hierarchy of simple ODEs for the $Q_i$, which can be solved analytically in series. Solving the equations that arise at $\order{1}$, $\order{\varepsilon}$, and $\order{\varepsilon^2}$ gives, respectively, $
Q_1(Y) = K_1 Y,  Q_2(Y) = K_2 Y^2,  Q_3(Y) =K_3 Y^3$, where the $K_i$, $i = 1,~2,~3$ are known functions of $\kappa$, $U_{\text{out}}$, $P_T$, and $Z_b$.

%where
%\begin{align}
%K_1 &=  Z_b'(X_p) U_\out,\\
%K_2 &= \frac{1}{6K_1^2}\left\{4\left[Z_b'(X_p)\right]^2 Z_b''(X_p)U_\out^3 \right. + \nonumber \\ \qquad \qquad &\left. \qquad  \kappa \left[Z_b'(X_p)\right]^4 \left[(1 - Z_b(X_p) - 2\Pt\right] K_1 - \left[Z_b'(X_p)\right] Q_\in\right\}\\
%K_3 &= \frac{\hat{K}_3 - 3K_1 K_2^2}{9K_1^2}.
%\end{align}
%where
%\begin{multline}
%        \hat{K}_3  = \kappa Z_b'(X_p)^4\left\{\left[1 - Z_b(X_p) - 2P_T \right]K_2 - Z_b'(X_p)K_1 - \frac{Z_b''(X_p)Q_\in}{2}\right\} + \\
% 4 \kappa  Z_b'(X_p)^3 Z_b''(X_p)\left\{\left[1 - Z_b(X_p) - 2P_T\right]K_1 - Z_b'(X_p)Q_\in\right\} + \\
% 2U_\out^3\left[Z_b'(X_p)^2 Z_b'''(X_p) + 3Z_b'(X_p) Z_b''(X_p)^2 \right].
%\end{multline}
%with $\hat{K}_3$ a known function of $Z_b$ and its derivatives at $X_p$, $P_T$, $\kappa$, and $U_\text{out}$.

Using $Q' = U Z_b'$ from before, the velocity associated with~\eqref{E:MeltRate:Q_expansionR3} is
 \begin{equation}\label{E:MeltRate:U_expansionR3}
U = U_{3l} \coloneqq \frac{1}{Z_b'(X)} \sum_{i = 1}^{\infty} \varepsilon^{i-1} \dd{Q_i}{Y}.
\end{equation}

We use~\eqref{E:MeltRate:U_expansionR3} to determine $X^*$: by retaining only the first three terms in the expansion~\eqref{E:MeltRate:U_expansionR3}, and making the approximation  $Z_b'(X) \approx Z_b'(X_p)$ (i.e. ignoring any variation in the ice shelf base in this region), we find that $X^*$\blue{, provided that it is less than $X_{\text{fr}}$,} must satisfy the quadratic equation
\begin{equation}\label{E:MeltRate:quadratic_for_xstar}
f U_\out = \frac{1}{Z_b'(X_p)}\left[K_1 + 2 K_2(X^* - X_p) + 3K_3(X^* - X_p)^2 \right].
\end{equation}
In what follows we take $f = 0.7$, i.e. $X^*$ is the \blue{maximum of $X_{\text{fr}}$} and the point at which the plume speed according to~\eqref{E:MeltRate:U_expansionR3} drops to approximately 70\% of $U_\out$. In the case that~\eqref{E:MeltRate:quadratic_for_xstar} has no solution, we assume that the plume reaches the ice shelf front without terminating, taking $X^* = X_{\text{fr}}$ and ignoring any further contributions to the melt rate. \blue{We verified that our constructed parametrization is insensitive to the choice of $f$, provided that both $f$ and $(1-f)$ remain $\order{1}$, i.e. region three is not dominated by either the lower or upper part.}

Having constructed an approximation to both the flux and velocity in the lower part of region three, we have the ingredients necessary to determine the thermal driving from~\eqref{E:Region3:thermal} and thus an approximation to the melt rate. However, since the expansion~\eqref{E:MeltRate:Q_expansionR3} only accounts for the ice shelf geometry at $X = X_p$, we postulate that it will perform poorly when applied to ice shelves with significant geometric variations above the pycnocline. In lieu of an analytic method that accounts for non-constant basal slopes, we apply the same ad-hoc geometric dependence as Lazeroms et al.~\cite{Lazeroms2018TheCryo}. Our approximation to melt rate therefore takes values
\begin{equation}\label{E:MeltRate:regions3_l}
M_{p}(X) = M_{p,3l}(X)\coloneqq Z_b'(X)^{5/2}\left\{\left[1  - 2\Pt -  Z_b(X)\right] U_{3l}(X) - Q_{3l}(X)\right\}
\end{equation}
for $X_p + N_l \lt < X < X^*$. [Explicitly, the prefactor $Z_b'(X)^{5/2}$ arises as the product of $Z_b'(X)$, which appears in the conservation of thermal driving~\eqref{E:Region3:thermal}, and $Z_b'(X)^{3/2}$, the scaling of the aspect ratio in the non-dimensionalization.]

Our construction in the upper part of region three, $X^* < X < X_c$, is motivated by the asymptotic behaviour as $X \to X_c$. Since $Q_c$, and thus the prefactor in~\eqref{E:Region3:UD_asym}, are unknown, we first express the velocity in $X^* < X < X_c$ according to the asymptotic behaviour as $X\to X_c$~\eqref{E:Region3:MatchingConditions}, albeit with an arbitrary prefactor:
\begin{equation}\label{E:MeltRate:Region3u_U}
U =U_{3u} =  C (X_c - X)^{1/3},
\end{equation} 
where $C$ and $X_c$ are to be determined. Note that since the solution in region four can be approximated by a solution of the form~\eqref{E:MeltRate:Region3u_U} [see \S3\ref{S:Asymptotics:Region4}], the below is also considered to be appropriate for region four.

After asserting that the velocity must be continuously differentiable across $X^*$, $C$ and $X_c$ are uniquely determined from~\eqref{E:MeltRate:U_expansionR3} and~\eqref{E:MeltRate:Region3u_U} as
\begin{equation}\label{E:MeltRate:Xc_expression}
X_c = X^* + \frac{U_{3l}(X^*)}{3 U_{3l}'(X^*)}, \qquad C = \frac{U_{3l}(X^*)}{\left(X_c - X^*\right)^{1/3}}
\end{equation}
In the case that this procedure gives $X_c < X^*$ [i.e. if $U_{3l}'(X_c) > 0$], we set $X^* = X_{\text{fr}}$ and the upper part of region three is considered moot.

Our approximation to the melt rate in $X^* < X <X_c$ is then constructed using~\eqref{E:Region3:thermal} and applying the ad-hoc geometric dependence:
\begin{equation}\label{E:MeltRate:regions3_u}
M_p(X) = M_{p,3u}(X) \coloneqq Z_b'(X)^{5/2}\left\{ \left[1  - 2\Pt -  Z_b(X)\right] U_{3u}(X) - Q_{3l}(X^*) \right\}
\end{equation}
for $X^* < X < X_c$. Note that in~\eqref{E:MeltRate:regions3_u}, we have made the approximation $Q \approx Q_{3l}(X^*)$ for all $X^* < X < X_c$, which is justified based on the fact that the flux is constant as $X \to X_c$ (see \S2 of the supplementary information).

In summary, except for the special cases mentioned above, our approximation to the dimensionless melt rate takes values
\begin{equation}\label{E:MeltRate:AllRegions}
M_{p}(X) = \begin{cases} 
M_{p,1}(X) \text{~~\quad [equation~\eqref{E:MeltRate:region1}]}  & 0 < X \leq X_p - N_l \lt,\\
M_{p,2}(X) \text{~~\quad [equation~\eqref{E:MeltRate:regions2}]} & X_p - N_l \lt < X \leq X_p + N_l \lt,\\
M_{p,3l}(X) \text{~\quad [equation~\eqref{E:MeltRate:regions3_l}]} & X_p + N_l \lt < X \leq X^*,\\
M_{p,3u}(X) \text{\quad [equation~\eqref{E:MeltRate:regions3_u}]} & X^* < X \leq X_c.
\end{cases}
\end{equation}
The corresponding construction in the special cases is set out explicitly in the supplementary information.

Our approximation to the dimensional melt rate is obtained by undoing the various scalings:
\begin{equation}\label{E:MeltRate:DimensionalMelt}
\dot{m} =\left(\frac{\beta_S S_l g  E_0^3 \alpha^3}{\lambda C_d (L/c)^3}\right)^{1/2} \tau^2  M_p\left(\frac{\lambda \alpha X}{\tau}\right),
\end{equation}
where the $X$ in the argument of $M_p$ is dimensional. We henceforth refer to~\eqref{E:MeltRate:DimensionalMelt} as the `B22 approximation'.

\section{Performance of approximation}\label{S:Numerics}
In this section, we assess the performance of the approximation~\eqref{E:MeltRate:DimensionalMelt} by comparing it to the melt rate obtained by numerically solving the full equations~\eqref{E:Modelling:Model:Mass}--\eqref{E:Modelling:Model:Thermal} \blue{(we henceforth refer to such numerical solutions as the `stratified plume model' for brevity)}, as well as to both the L19 and L19AH approximations. Assessment of the performance of the B22 approximation is both based upon agreement with \rout{numerical solutions }\blue{the stratified plume model} and improving upon the L19 and L19AH approximations. \blue{In addition, it is desiable that the B22 approximation be computationally less expensive than the stratified plume model. This is indeed the case: for the examples presented here, the B22 approximation typically requires a computation time that is an order of magnitude smaller than that required by the stratified plume model.} Note that the comparisons presented in this section use the dimensional form for consistency with the literature on basal melt rates under ice shelves.

In theory, we have made two improvements upon the L19 and L19AH approximations: firstly, \blue{an improved consideration of ice shelf basal geometry, which accounts for the effect of the upstream, as well as local, geometry}\rout{accounting for a general ice shelf basal geometry and capturing non-local geometric effects}, and, secondly, accounting for stratification in the ambient salinity and temperature. In the first part of this section, we consider both of these improvements in isolation: we begin by assessing the performance of the B22 approximation for different ice shelf basal geometries with constant ambient conditions. Following this, we assess its performance on an ice shelf with a constant basal slope for various pycnocline positions \blue{and thicknesses}. Finally, we combine these two, and assess the performance of the B22 approximation in geometries with a non-constant basal slope and ambient stratification.

\subsection{Isolating Basal Geometry}\label{S:Numerics:NoPycnocline}
\begin{figure}
\centering
\includegraphics[width = 0.9\textwidth]{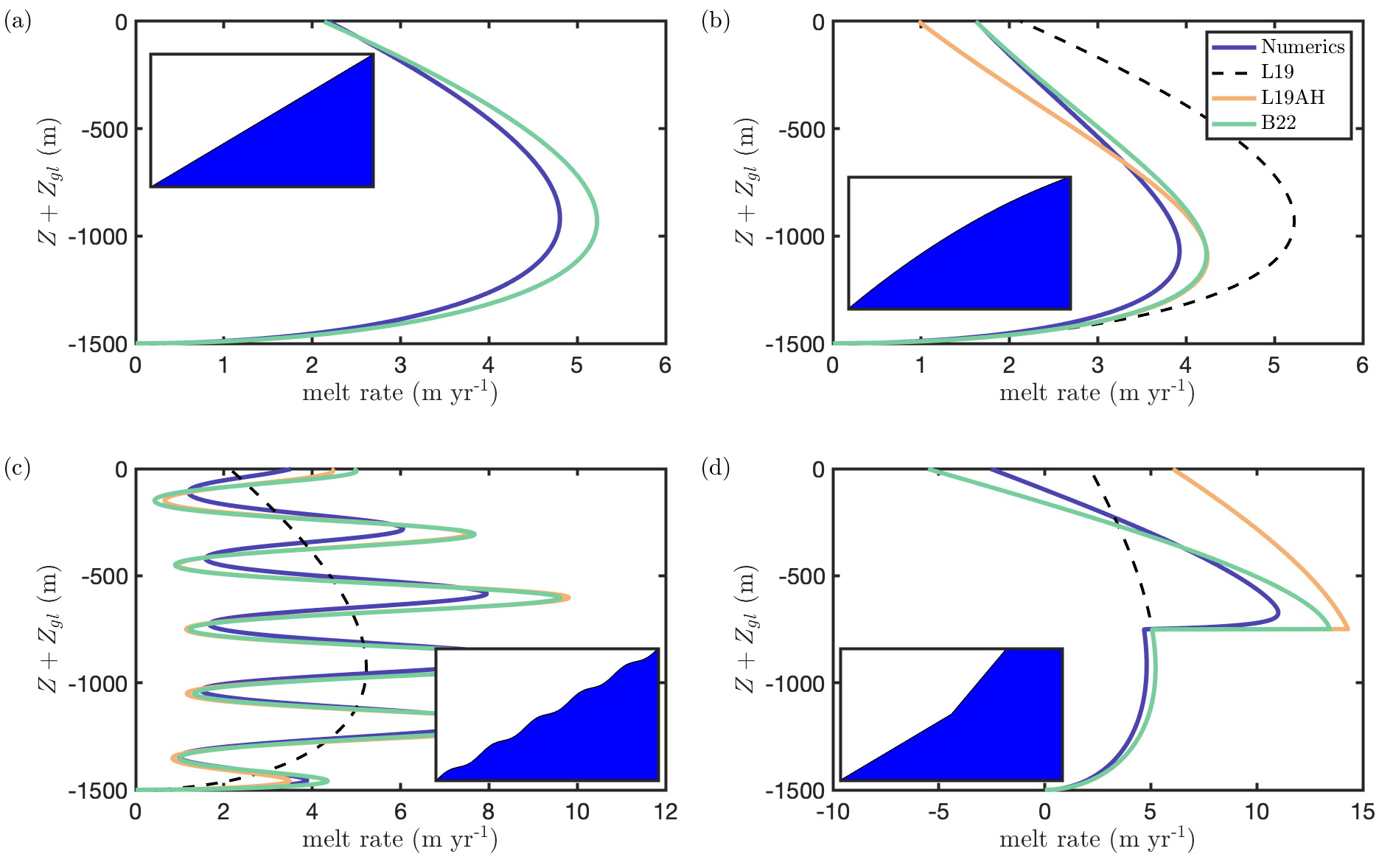}
\caption{Comparison between\rout{ numerically obtained} melt rates $\dot{m}= M_0 U \Delta T$ \blue{from the stratified plume model} (purple curves) and the three approximations: B22 (green curves), L19AH (orange curves), and L19 (black dashed curves). Each subplot corresponds to a different ice shelf basal geometry as follows (shown schematically in the insets): (a) linear, $Z_b(X) = \alpha X$, (b) quadratic~\eqref{E:Numerics:QuadraticGeometry}, (c) linear with periodic oscillations superimposed~\eqref{E:Numerics:SinusoidalGeometry}, and (d) piecewise linear~\eqref{E:Numerics:PiecewiseGeometry}. Note that the three approximations to the melt rate are identical for the constant basal slope case in (a).  Here we use parameter values as in table~\ref{T:Constants}, except for imposing a constant ambient temperature and salinity, which we take as $ 0.5\si{\celsius}$ and $34.3~\text{PSU}$, respectively. Note that the upper limit of each the plots, $Z + Z_{gl} = 0$, corresponds to sea level, which acts as an upper bound on the extent of the ice shelf base.}   \label{fig:Numerics:Geometry}
\end{figure}

In figure~\ref{fig:Numerics:Geometry}, we present a comparison between \rout{numerically obtained melt rates}\blue{the melt rate from the stratified plume model} and the three approximations, for several basal geometries using constant ambient conditions. [For the B22 approximation, this unstratified case is recovered by setting $X_p = \infty$ in~\eqref{E:MeltRate:DimensionalMelt}.] The scenario with constant ambient conditions may also be pertinent if the pycnocline is below the level of the grounding line, or if an obstacle such as a seabed ridge obscures access of the warm, salty lower layer to the ice front, as is the case for Pine Island Glacier~\cite{DeRydt2014JGeophysResOceans}. 

%talk about each of the drafts in question]
In the case of a constant basal slope (figure~\ref{fig:Numerics:Geometry}a), the melt rate \blue{in the stratified plume model} again displays the characteristic \rout{subglacial plume} behaviour that was described in \S2\ref{S:ExampleSolutions}: briefly, the melt rate increases from the grounding line, reaching a maximum several hundreds of metres downstream, before decreasing towards zero. For the ambient temperature used to generate the solutions shown in figure~\ref{fig:Numerics:Geometry}a, the thermal driving is strong, and thus the melt rates remain positive: the ice shelf is not long enough for the plume to become supercooled. Each of the three approximations are identical in this constant slope case, and show good agreement with the \rout{numerical solution }\blue{stratified plume model} (figure~\ref{fig:Numerics:Geometry}a).

The pattern of melt rates is broadly similar for the quadratic profile (figure~\ref{fig:Numerics:Geometry}b) given by
\begin{equation}\label{E:Numerics:QuadraticGeometry}
\frac{Z_b(X)}{\left[Z\right]}= \frac{X}{\left[X\right]} - \frac{1}{2}\left(\frac{X}{\left[X\right]}\right)^2,
\end{equation}
albeit that the maximum melt rate is reduced, and the peak melt rate occurs closer to the grounding line. This geometry can be thought of as typical of one in which changes in the basal slope occur on the same length scale as the ice shelf. The B22 approximation agrees well with the\rout{ numerical solution} \blue{stratified plume model} throughout, while the L19AH parametrization agrees well with\rout{ numerical solutions} \blue{stratified plume model} close to the grounding line, but deviates closer to the ocean surface; errors in the L19AH approximation increase with distance from the grounding line, which is indicative of the accumulation of errors resulting from consideration only of local rather than all upstream geometry. Despite this, the L19AH approximation performs significantly better than the L19 parametrization, highlighting that the ad-hoc dependence used by the L19AH approximation is effective at capturing the effect of variations in the basal slope.

In figure~\ref{fig:Numerics:Geometry}c, we present the comparison for the case of a linear ice shelf base with periodic oscillations superimposed:
\begin{equation}\label{E:Numerics:SinusoidalGeometry}
\frac{Z_b(X)}{\left[Z\right]}= \frac{X}{\left[X\right]} + \frac{30}{\left[Z\right]}\sin\left(\frac{2\pi}{0.1}\frac{ X}{\left[X\right]}\right).
\end{equation}
Note that period of oscillation, and thus length scale on which basal slope changes occur, is relatively short in comparison with the length scale of the freezing point dependence (indicated by the $0.1$ in the denominator in the argument). Both the L19AH and B22 approximations accurately capture the oscillatory melt rate seen in \blue{the stratified plume model}\rout{ numerical solutions}, but overestimate the melt rate at troughs of the profile where the basal slope approaches zero; in these areas, our assumption that the slope is $\order{\alpha}$ breaks down, and the relatively good agreement is therefore somewhat surprising. Unlike in the quadratic case, the L19 approximation shows good agreement \blue{with the stratified plume model} throughout the water column; the errors that accumulate with depth effectively cancel out between peaks and troughs of the profile.  

Finally, in figure~\ref{fig:Numerics:Geometry}d, we present the comparison in the case of a piecewise linear ice shelf base,
\begin{equation}\label{E:Numerics:PiecewiseGeometry}
\frac{Z_b(X)}{\left[Z\right]} = \begin{cases}
  \frac{X}{\left[X\right]}  & 0 <   \frac{X}{\alpha |Z_{gl}|} \leq   \frac{1}{2},\\
   2 \left(\frac{X}{\left[X\right]} -\frac{1}{2}\right)  &  \frac{1}{2} <   \frac{X}{\alpha |Z_{gl}|} <   1.
    \end{cases}
\end{equation}
This profile features a doubling of the slope at a depth that is halfway between the grounding line and the sea surface. Below the discontinuity, the approximations are identical and the agreement with numerical solutions is very good. Both the L19AH and B22 approximations significantly overestimate the melt rate in the region around the discontinuity, where, again, our assumption that the basal slope is $\order{\alpha}$ breaks down. The L19 approximation does not have any information about the change in geometry and thus significantly underestimates buoyancy driving and thus melt rates there. Above the discontinuity, the three approximations deviate, but the B22 approximation displays the best agreement with the \blue{stratified plume model}\rout{ numerical solution}. It is also worth noting that the performance of the B22 approximation is superior to the L19AH approximation in the region just beyond the discontinuity, indicating that the former is better able to capturing the local, in addition to global, dependence on the geometry.

In summary, in the absence of stratification the approximation constructed in \S\ref{S:MeltRate} represents an improvement on both the L19 and L19AH approximations, on account of its ability to better capture variations in the slope of the ice shelf base both locally and non-locally. In addition, provided that the assumptions on the geometry hold, \blue{the approximation constructed in \S\ref{S:MeltRate} is exactly correct to leading order [$\order{1}$]; the next term in an asymptotic expansion of the equations appropriate to region one, and thus the error introduced by considering only the leading order behaviour, is $\order{\epsone}$} \rout{we have formally quantified the error as}\blue{in the limit $\epsone \to 0$}; no such \blue{formally} constrained error exists for either the L19 or L19AH approximations. 

\subsection{Isolating Stratification}
\begin{figure}
\centering
\includegraphics[width = \textwidth]{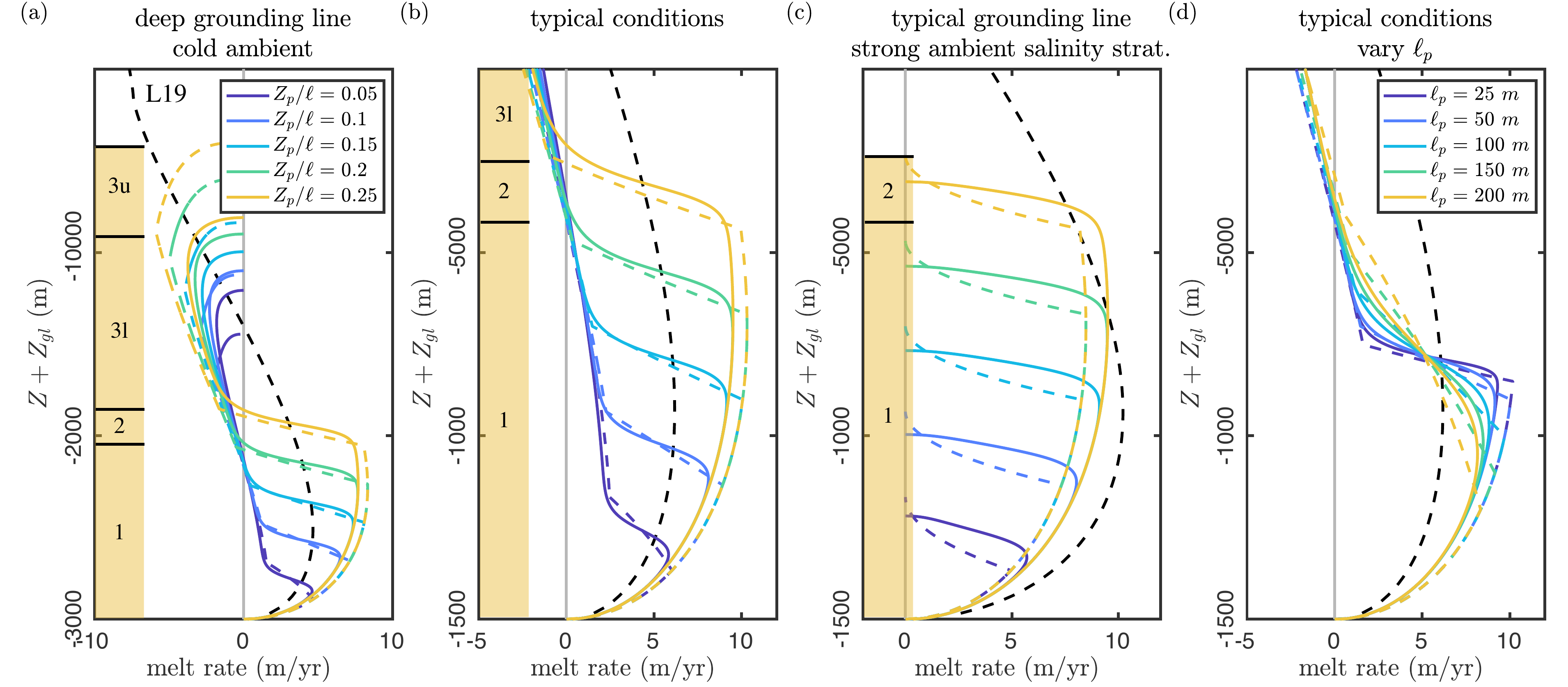}
\caption{Comparison between \rout{numerically obtained }melt rates $\dot{m}= M_0 U \Delta T$ \blue{from the stratified plume model} (solid curves), the B22 approximation (coloured dashed curves) and the L19 approximation (black dashed curve), for the case of two layer stratification and a constant basal slope. \blue{In (a)--(c),} \rout{R}\blue{r}esults are shown for different pycnocline positions $Z_p$ [colours, as indicated by the legend in (a)]. In each \rout{case}\blue{of these}, the parameters are as in table~\ref{T:Constants}, except for (a) $Z_{gl} = -3000$~m, $T_l = -1~\si{\celsius}$ and $T_u = -3~\si{\celsius}$\blue{, which correspond to a deep grounding line, and cold ambient conditions,} and (c)  $S_u = 33.0$~PSU\blue{, which corresponds to strong ambient salinity stratification}.  The panels along the left-hand side of each \rout{subplot}\blue{of (a)--(c)} indicate which of the regions discussed in \S\ref{S:MeltRate} for the B22 approximation is appropriate at that particular depth, for the case of a pycnocline centred at $Z_p/\ell = 0.25$. \blue{In (d), results are shown for different pycnocline thicknesses $\ell_p$ (colours, as indicated by the legend). All parameters are as in table~\ref{T:Constants}, with the pycnocline centred on a depth $Z = 500$~m.} Note that for the L19 approximation, we take the constant ambient temperature and salinity to be the average of the values above and below the pycnocline \blue{centre}.}\label{fig:Numerics:PycnoclinePosition}
\end{figure}

In figure~\ref{fig:Numerics:PycnoclinePosition}, we present a comparison between \blue{melt rates from the stratified plume model}\rout{numerically obtained melt rates} and the approximations for the case of a constant basal slope and two layer stratification. This comparison allows us to isolate the effect of ambient stratification from geometry on differences between the approximations and \blue{the stratified plume model}\rout{numerical solutions}. \blue{Within this, we first consider different pycnocline positions for various combinations of ambient conditions and grounding line depth (figure~\ref{fig:Numerics:PycnoclinePosition}a--c), and then consider various pycnocline thicknesses (figure~\ref{fig:Numerics:PycnoclinePosition}d).}  Note that in the case of a constant basal slope the L19 and L19AH approximations are identical, and we therefore refer only to the former in this section, and use the average value of the upper and lower layer salinity and temperature as the respective constant values in this approximation.

The parameters used in figure~\ref{fig:Numerics:PycnoclinePosition}\blue{a--c} represent three different scenarios: (a) a deep grounding line with cold ambient conditions and typical ambient salinity; (b) typical grounding line depth and ambient conditions; and (c) typical grounding line depth and ambient temperature, but strong salinity stratification \blue{(specific values of those parameters used to generate results shown in figure~\ref{fig:Numerics:PycnoclinePosition} are set out in the caption of this figure)}.

%comparison numerics and my melt only (a)
For cold, deep conditions (figure~\ref{fig:Numerics:PycnoclinePosition}a) we see very good agreement between \blue{the stratified plume model}\rout{numerically obtained melt rates} and the B22 approximation up until well beyond the pycnocline. There, the B22 approximation captures the broad trend of the \blue{stratified plume model}\rout{numerical solutions}, characterized by a strong refreezing before a sharp transition to zero melt when the plume terminates, but typically over-estimates the peak refreezing rate. The loss of accuracy at shallower depths (larger $Z + Z_{gl}$) is concomitant with errors that increase with distance from the pycnocline, as is to be expected from having expanded about this point.

%comparison numerics and my melt only (b)
For a more typical grounding line depth and temperature profile (figure~\ref{fig:Numerics:PycnoclinePosition}b), the agreement between the B22 approximation and \blue{the stratified plume model}\rout{numerical solutions} is very good everywhere. In contrast to the cold, deep case, the plume reaches the surface having not spent a significant distance in region three, and the errors associated with the expansion~\eqref{E:MeltRate:U_expansionR3} are thus relatively small.

\rout{Finally, f}\blue{F}or the case of unusually strong ambient salinity stratification (figure~\ref{fig:Numerics:PycnoclinePosition}c), the B22 approximation is able to accurately capture the intrusion of the  plume into the ambient. Agreement between the B22 approximation and \blue{the stratified plume model}\rout{numerical solutions} is again very good everywhere and, in particular, the B22 approximation accurately predicts the point at which the plume intrudes into the ambient.

\blue{In figure~\ref{fig:Numerics:PycnoclinePosition}d, we present the comparison for different pycnocline thicknesses $\ell_p$. As the pycnocline thickness increases, as does the dimensionless parameter $\delta$, and we therefore expect the accuracy of solutions to be reduced. This is borne out in the solutions: for a very narrow pycnocline ($\ell_p = 25$~m, darkest curve) agreement is excellent throughout the water column; as the pycnocline thickness increases, the accuracy of solutions reduces, with the melt rate increasingly underestimated below the pycnocline, and overestimated above the pycnocline. Despite this, even with a pycnocline half length of $\ell_p = 200$~m, a reasonable upper bound on realistic conditions, agreement between our approximation and the stratified plume model is still good throughout the water column. This is primarily because the lengthscale on which the freezing point changes with depth is on the order of kilometres, while the lengthscale of the pycnocline is on the order of hundreds of meters: the parameter $\lt$ thus remains relatively small for all values of $\ell_p$ up to this upper bound.}

Figure~\ref{fig:Numerics:PycnoclinePosition} demonstrates that the B22 approximation represents an improvement over the L19 approximation in the case that the ambient ocean has two layer stratification. The latter parametrization -- which, it should be stressed, is only designed to account for constant ambient conditions -- fails to capture the behaviour \blue{of the stratified plume model} above the pycnocline: the L19 approximation significantly over-predicts the portion of the ice shelf base in which melting occurs and fails to capture the transition to refreezing.

\subsection{Including Non-planar Basal Geometry and Stratification}
\begin{figure}
\centering
\includegraphics[width = \textwidth]{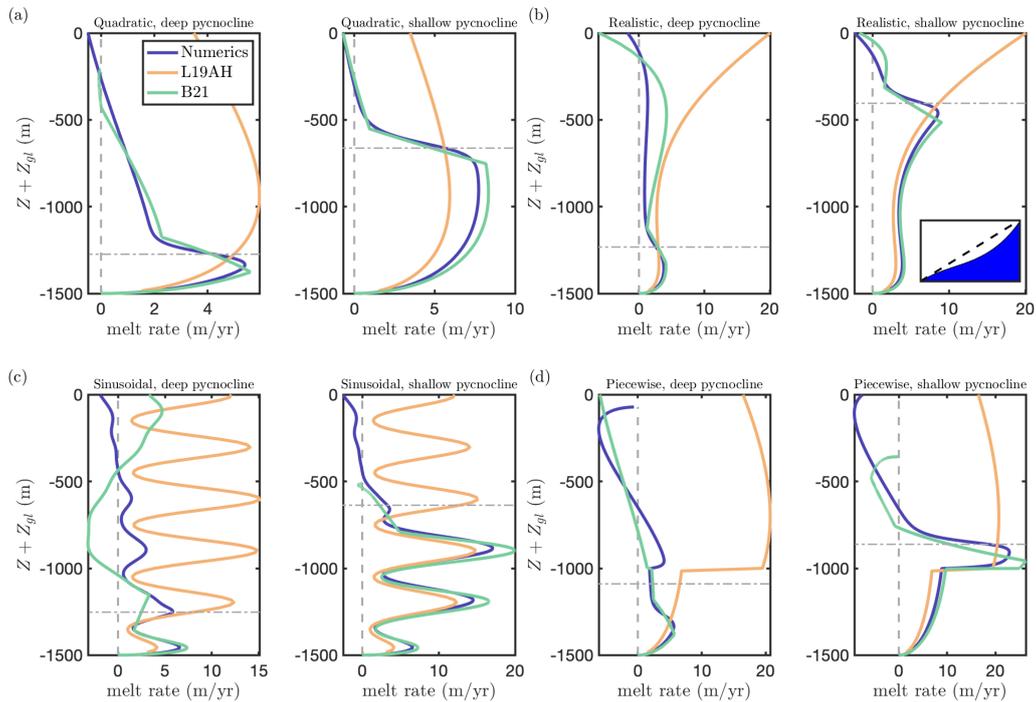}
\caption{Comparison between \rout{numerically obtained }melt rates $\dot{m}= M_0 U \Delta T$ \blue{from the stratified plume model} (purple curves), the B22 approximation (green curves) and the L19AH approximation (orange curves), with two layer stratification. Results are shown for several different basal geometries as follows: (a) `quadratic'~\eqref{E:Numerics:QuadraticGeometry}, (b) `idealized Ross'~\eqref{E:Numerics:RealisticGeometry}, (c) `sinusoidal'~\eqref{E:Numerics:SinusoidalGeometry}, and (d) `piecewise linear' ~\eqref{E:Numerics:PiecewiseGeometry} (the geometries corresponding to (a), (b) and (d) are shown schematically in the insets in figure~\ref{fig:Numerics:Geometry}(b), (c) and (d), respectively, while the geometry corresponding to (b) is shown as an inset in that plot). In each subplot, the left and right hand plots correspond a relatively deep pycnocline and relatively shallow pycnocline, respectively \blue{; the pycnocline position is indicated by the grey dot-dashed line}. The parameters used to generate the results shown here are as in table~\ref{T:Constants}. \blue{Vertical grey dashed lines indicate the a zero melt rate.} }\label{fig:Numerics:pycnocline_and_geometry_idealized}
\end{figure}

Finally, we assess the performance of the B22 approximation in the case of a non-constant basal slope and two-layer ambient stratification. The comparison between the approximations and \blue{the stratified plume model} \rout{numerical solutions} is presented in figure~\ref{fig:Numerics:pycnocline_and_geometry_idealized} [we do not include the L19 approximation in this assessment, as it was shown in \S5\ref{S:Numerics:NoPycnocline} to perform poorly for non-constant basal slopes]. Here, we consider two different pycnocline positions, one relatively deep and one relatively shallow, for each of four geometries: the quadratic, sinusoidal, and piecewise linear geometries referred to earlier [shown schematically in the insets in figure~\ref{fig:Numerics:Geometry}(b), (c), and (d), respectively], as well as \rout{a fourth geometry} \blue{an additional geometry} 
\begin{equation}\label{E:Numerics:RealisticGeometry}
\frac{Z_b(X)}{\left[Z\right]}= \frac{X}{\left[X\right]} -4.2\left(\frac{X}{\left[X\right]}\right)^2 + 12.8\left(\frac{X}{\left[X\right]}\right)^3.
\end{equation}
We refer to the geometry~\eqref{E:Numerics:RealisticGeometry} as `idealized Ross', owing to its qualitative similarity to a transect taken along a flowline under the Ross Ice Shelf~\cite{Shabtaie1987JGeophysResSolidEarth}. In particular, this geometry features a flatter section in the centre of the ice shelf base and a relatively large positive slope close to its front (see inset in figure~\ref{fig:Numerics:pycnocline_and_geometry_idealized}b).

In both the quadratic and the idealized Ross cases, the salient observations from the comparison are, firstly, that agreement between \rout{numerical solutions}\blue{the stratified plume model} and the B22 approximation is very good and, secondly, that the B22 approximation performs significantly better than the L19AH approximation. Crucially, in these cases, the shape of the ice shelf base satisfies the assumptions made in the asymptotic analysis: their slopes remain $\order{\alpha}$ throughout, with well-defined derivatives which vary on the same scale as the length of the base. In the sinusoidal and piecewise cases, where these assumptions are violated, however, large differences between numerical solutions and the B22 approximation are observed.

In more detail, the pattern of melting in the quadratic case (figure~\ref{fig:Numerics:pycnocline_and_geometry_idealized}a) with two layer stratification is very similar to that in the corresponding linear case (figure~\ref{fig:Numerics:PycnoclinePosition}): a melt rate that increases with distance from the grounding line to a peak located just below the pycnocline if the pycnocline is deep (figure~\ref{fig:Numerics:pycnocline_and_geometry_idealized}a left), and some distance below the pycnocline if it is shallow (right), with refreezing close to the ice shelf front. These features are all captured well by the B22 approximation. The agreement above the pycnocline reinforces our assertion that the ad-hoc dependence used by the B22 approximation accounts well for geometric variations, provided that the geometry is not overly complex. Although agreement is good for both pycnocline positions in the idealized Ross case (figure~\ref{fig:Numerics:pycnocline_and_geometry_idealized}b), it is somewhat better in the shallow pycnocline case. In the deep case, the geometry downstream of the pycnocline is relatively complex; the over-prediction of both melting and refreezing rates above the pycnocline (figure~\ref{fig:Numerics:pycnocline_and_geometry_idealized}b, left) are a consequence of both increasing errors with distance from the pycnocline, and the inability of the ad-hoc method to completely capture the geometric dependence in this case. 

The sinusoidal case (figure~\ref{fig:Numerics:pycnocline_and_geometry_idealized}c) demonstrates clearly that the B22 approximation may perform poorly when the geometry does not conform to the restrictions placed on it. Although agreement is good below the pycnocline [as discussed in \S5\ref{S:Numerics:NoPycnocline}], this parametrization makes spurious predictions through and above the pycnocline. There are two main reasons for this: firstly, the upper edge of the discrete pycnocline region introduced in \S\ref{S:MeltRate} is located near a trough of the profile, where the slope approaches zero, resulting in large errors in the expansion~\eqref{E:MeltRate:U_expansionR3} (which has the basal slope in its denominator).  Secondly, the slope of the base changes significantly across the pycnocline, leading to an inaccurate representation of the important physics in this region (we assumed, when analyzing the pycnocline region in \S\ref{S:Asymptotics}, that the ice shelf basal slope does not change significantly within it).

Finally, in the piecewise linear case (figure~\ref{fig:Numerics:pycnocline_and_geometry_idealized}d), the agreement between \blue{the stratified plume model}\rout{numerical solutions} and the B22 approximation is surprisingly good for the most part, although there is a significant error for the plume intrusion depth in the high pycnocline case. This agreement occurs despite the violation of the geometric assumptions, and is especially surprising for a low pycnocline: the discontinuity in the geometry occurs above the pycnocline, where we expect errors associated with the geometry to be accentuated. 

\section{Discussion and Conclusions}\label{S:Discussion}
We have presented an asymptotic analysis of the equations describing subglacial plumes~\cite{Jenkins1991JGeophysResOceans, Jenkins2011JPhysOcean} with non-constant basal slopes and two layer ambient stratification. We then constructed an approximation to the melt rate that emerges from numerical solutions of the stratified plume model equations, which builds upon this analysis. This approximation agrees well with \blue{the stratified plume model}\rout{numerical solutions}, provided that assumptions on the basal topography are not violated. We further demonstrated that this approximation improves upon that of Lazeroms et al.~\cite{Lazeroms2019JPhysOcean}, on which some of the current state of the art two-dimensional melt rate parametrizations are based.

Our asymptotic analysis exploits the small size of several dimensionless parameters; in particular, the small basal slopes that are typical for ice shelves in Antarctica, the relative slenderness of the pycnocline, and the relatively weak thermal driving. This analysis reveals the presence of four distinct solution regions, depending on the distance to the grounding line, the distance to the pycnocline, and the speed of the plume. The approximation to the melt rate we constructed was similarly split according to these regions, albeit with clear boundaries between them rather than matching conditions, as in the asymptotic analysis. Only in the region closest to the grounding line do the corresponding leading order equations have an analytic solution, i.e. our approximation to the melt rate has an asymptotically well-defined error only in this region. 

Another way to view this is that, in the absence of stratification, or in the case that the level of pycnocline sits below the grounding line, our approximation to the melt rate has a small, well-defined error everywhere. This was demonstrated by comparison with numerical solutions: the approximation we constructed agrees well for a variety of idealized ice shelf bases, even when the geometric assumptions made during its derivation were violated. In addition, it performs better than the Lazeroms et al.~\cite{Lazeroms2019JPhysOcean} approximation, which was attributed to an improved accounting of both local and non-local geometry. \blue{The ambient temperature and salinity profiles we chose in this study are motviated by typical conditions in the Amundsen Sea, where a large flux of CDW is able to spill onto the continental shelf. Elsewhere around Antarctica, little CDW is able to spill onto the continental shelf and the water column is therefore less stratified~\cite{Petty2013JPO}; thus, we suggest that the analysis presented here, which represents an improvement on previous studies with an unstratified ambient, is also applicable to regions of Antarctica outside of the Amundsen Sea Sector.}

When ambient stratification is imposed on the model, further solution regions must be considered. When this stratification is strong, the plume may intrude into the ambient within the pycnocline. In this case, only two of the four regions considered are pertinent; our approximation accurately predicted both the depth at which the plume terminates and the behaviour up to this point. \blue{The stratification used to generate these results is stronger than is observed in Antarctica, supporting the suggestion that plume intrusion into the ambient}\rout{ This behaviour} is only expected when stratification is stronger than is typical for oceans surrounding Antarctica\blue{. These highly stratified ambient conditions may be appropriate for Greenland, and so we speculate  that the approximation derived here may also be applicable to outlet glaciers located therein.} With more typical Antarctic conditions, each of the four regions enters into the solution, and our approximation also showed good agreement with numerical solutions in this case\blue{, even when the pycnocline thickness was varied up to a value that might be considered an upper bound of what is observed in practice}. We saw a reasonable deviation, however, when the grounding line is deep, in which case the plume terminates before it reaches the ice shelf front, which is attributed to several approximations made when describing the behaviour above the pycnocline in our construction.

Finally, we assessed the performance of our approximation in the case that both two-layer ambient stratification and a non-constant basal slope are imposed. Good agreement was seen in cases in which the assumptions on the basal topography were not violated, but this comparison also highlighted the possibility of spurious predictions when these assumptions on the geometry do not hold.

The main motivation behind this study is improving two-dimensional parametrizations of melt rates for use in future projections of the Antarctic ice sheet. The key challenge, following the analysis presented here, is to extend the constructed approximation to a second spatial dimension. \rout{Whilst this is beyond the scope of this paper, it should be noted that this presents a significant challenge.} Lazeroms et al.~\cite{Lazeroms2018TheCryo} discussed this issue in detail and offered a (non-unique) solution; unfortunately, the method described therein is not immediately applicable to the approximation constructed here since it requires that the approximation to the one-dimensional plume theory melt rate depends only on the local ice shelf conditions, whereas the approximation constructed here requires the history of the plume path to be resolved. \blue{The challenge, therefore, is perhaps how to define the plume path. One way to achieve this might be to determine, for a given grid point in an ice shelf, each direction in which the plume might possibly reach the point (as in Lazeroms et al.~\cite{Lazeroms2018TheCryo}), and then construct a plume path by extending a linear section from this grid point to the grounding line. The ice shelf geometry would then given by the vertical slice along this line. }

\rout{Other possible pragmatic} \blue{There are also potential} issues \blue{arising from our modelling assumptions. One such assumption is}\rout{include} our use of an idealized ambient stratification, which is only a first approximation to realistic conditions\blue{. In particular, application of our parametrization requires us to determine the pycnocline centre, as well as various co-ordinates, such as the location of plume termination. One possibility is that a fit of the form of the ambient profiles is applied to the realistic profiles of temperature and salinity, from which the pycnocline position may be recovered; after doing so, and determining the flowline path, as suggested above, the values of other required co-ordinates can be determined according to the procedure described in \S\ref{S:MeltRate}.} \rout{, and} \blue{Innacuracies in a parametrization based on the approximation derived here might also arise from our neglect of frazil ice dynamics, which can significantly alter the behaviour of subglacial plumes in regions where the melt rate is negative~\cite{ReesJones2018Cryo}. Furthermore, }\rout{our neglect of}\blue{we have neglected} subglacial discharge; although \rout{the latter}\blue{subglacial discharge} is expected to have a negligible influence on the plume dynamics \blue{on the length scale of the entire ice shelf}, \blue{there is evidence that it is responsible for the high melt rates close to the grounding line of Pine Island Glacier~\cite{Nakayama2021GRL}. } \rout{it may permit a high melt rate near to the grounding line, which has been shown to have a significant influence on the dynamics of the entire ice sheet}\blue{This is especially pertinent because enhanced melt rates close to the grounding line may disproportionately impact the dynamics of the ice sheet~\cite{Arthern2017GRL,Seroussi2018Cryo}. We suggest that plume-based melt rate parametrizations might be modified to account for subglacial discharge by including another asymptotic region in proximity to the grounding line, on the lengthscale of which the role of subglacial discharge can be resolved.} \blue{Finally, there is significant uncertainty in many of the parameters used in the construction presented here, most notably perhaps the entrainment coefficient $E_0$ and drag coefficient $C_d$. In this paper, we have considered a single set of of parameter values; accounting for uncertainty in these parameters may lead to different asymptotic balances, which future studies might explore. However, we expect that the leading order behaviour, as considered here, will be qualitatively similar in these cases, provided that the small parameters remain small within these uncertainties. Parameter uncertainty also provides motivation for improving our understanding of small scale ice-ocean processes in order to better constrain these parameters, and thus improve melt rate parametrizations.}

There are further practical considerations for the work presented here. Most pressing is the use of one dimensional plume theory: for example, any parametrization that is based on an extension of this theory will not (naturally) include rotation, and will not account for mode two or mode three melting, which are prominent in certain regions of Antarctica~\cite{Adusumilli2020NatureGeo}. The importance of accurate basal melt rates in projections of the Antarctic ice sheets on multi-century timescales provides motivation for the development of parametrizations that address these issues. 

\enlargethispage{20pt}

%\dataccess{Insert data access text here.}

\aucontribute{AB and RA conceived of the study. AB produced the analysis and drafted the manuscript. RA and CRW helped draft the manuscript and supervised the project. AJ critically reviewed the manuscript.}

%\competing{Insert competing text here.}

%\funding{Insert funding text here.}

\funding{A. T. B. is supported by the NERC Grant NE/S010475/1.}

%%%%%%%%%% Insert bibliography here %%%%%%%%%%%%%%
\bibliographystyle{vancouver}

\bibliography{mybib.bib}

\end{document}

\maketitle
\newcommand{\Pb}{\textit{P}_B}  %\frac{L/c}{ \tau}\frac{S_l - S_u}{2S_l},
\newcommand{\lt}{\delta} %dimensionless thermocline length, lt/l0
\newcommand{\Pt}{\textit{P}_T}
\renewcommand{\in}{\text{in}} %subcript on into and out of pycnocline variables
\newcommand{\out}{\text{out}}
\newcommand{\order}[1]{\mathcal{O}(#1)}
 \newcommand{\dd}[2]{\frac{\mathrm{d} #1}{\mathrm{d} #2}}
 \newcommand{\epsone}{\epsilon_1}

In this supplementary information we provides further details on the analysis presented in the main text. This suppelementary consists of five sections: in the first, We explicitly state the solution in region one and associated matching conditions for region two; in the second, we explicitly state the pycnocline-rescaled plume model equations; in the third, we provide details of the behaviour of solutions in region three in the limit $X \to X_c$; in the fourth section we provide further details of the behaviour in region 4; finally, we explicitly state our constructed melt rate parametrizations in special cases discussed in \S4 of the main text.

\section{Solution for Region One and Matching Conditions on Region Two}
The solution to the leading order equations appropriate for region one, given by (3.1)--(3.4) in the main text, is:
\begin{align}
U(X) &=\left(\frac{2\kappa}{3}\right)^{1/2}\left[Z_b'(X)\right]^{1/3}\left[1 - Z_b(X)\right]^{1/3}I(X)^{1/2},\\
%D &= \frac{2}{3}\left[Z_b'(X)\right]^{-1/3}\left[1 - Z_b(X)\right]^{-1/3}I(X),\label{E:Region1:thickness_solution}\\
D(X) &= \frac{2}{3}\left[Z_b'(X)\right]^{-1/3}\left[1 - Z_b(X)\right]^{-1/3}I(X),\label{E:Region1:thickness_solution}\\
\Delta \rho(X) &= \kappa \left[ 1- Z_b(X)\right],\label{E:Region1:buoyancy_solution}\\
%\Delta T &= Z_b'(X)\left[1 - Z_b(X)\right]\left\{1 - %\frac{2}{3}I(X)\left[Z_b'(X)\right]^{1/3}\left[1-Z_b(X)\right]^{-4/3}\right\}
\Delta T &= \left[1-Z_b(X)\right]Z_b'(X) - \frac{2}{3}\left[Z_b'(X)\right]^{2/3} \left[1-Z_b(X)\right]^{-1/3}I(X),\label{E:Region1:thermal_solution}
\end{align}
where
\begin{equation}
I(X) =  \int_0^X \left[Z_b'(\xi)\right]^{4/3}\left[1 - Z_b(\xi)\right]^{1/3}~\mathrm{d}\xi.
\end{equation}
The matching conditions on region two are therefore given by
\begin{align}
U \to U_\text{in} &=\left(\frac{2\kappa}{3}\right)^{1/2}\left[Z_b'(X_p)\right]^{1/3}\left[1 - Z_b(X_p)\right]^{1/3}I(X)^{1/2},\\
%D &= \frac{2}{3}\left[Z_b'(X)\right]^{-1/3}\left[1 - Z_b(X)\right]^{-1/3}I(X),\label{E:Region1:thickness_solution}\\
D \to D_\text{in} &= \frac{2}{3}\left[Z_b'(X_p)\right]^{-1/3}\left[1 - Z_b(X_p)\right]^{-1/3}I(X_p),\label{E:Region1:thickness_solution}\\
\Delta \rho \to \Delta \rho_\text{in} &= \kappa \left[ 1- Z_b(X_p)\right],\label{E:Region1:buoyancy_solution}\\
%\Delta T &= Z_b'(X)\left[1 - Z_b(X)\right]\left\{1 - %\frac{2}{3}I(X)\left[Z_b'(X)\right]^{1/3}\left[1-Z_b(X)\right]^{-4/3}\right\}
\Delta T \to \Delta T_{\text{in}} &= \left[1-Z_b(X_p)\right]Z_b'(X_p) - \frac{2}{3}\left[Z_b'(X)\right]^{2/3} \left[1-Z_b(X)\right]^{-1/3}I(X),
\end{align}
as $\zeta = (X - X_p)/\delta \to -\infty$.

\section{Pycnocline-rescaled plume model equations}
The pycnocline-rescaled plume model equations referred to in \S3.2 in the main text read:
\begin{align}
\dd{(DU)}{\zeta} &=\lt U Z_b'(X_p + \lt \zeta) + k_3 \epsone \lt U \Delta T,		\label{E:PycnoclineRegion:MassFull}	\\
\cone \dd{(DU^2)}{\zeta} &=  D\Delta \rho Z_b'(X_p + \lt \zeta)  -  U^2,	\label{E:PycnoclineRegion:MomFull}	\\
\dd{(DU\Delta \rho)}{\zeta} &= -\Pb \mathrm{sech}^2(\zeta)Z_b'(X_p + \lt \zeta)DU + \lt \left[\kappa - k_4 \epsone \tanh(\zeta)\right] U \Delta T \label{E:PycnoclineRegion:BuoyancyFull}		\\
\ctwo \dd{(DU\Delta T)}{\zeta} &= \left\{1 - Z_b(X_p + \lt \zeta) - \Pt\left[1 + \tanh(\zeta)\right] - D\right\}UZ_b'(X_p + \lt \zeta) -U\Delta T.\label{E:PycnoclineRegion:ThermalFull}
\end{align}

%%%%%%%%%%%%%%%%%%%%%% Analysis of Region 3 as X \to X_c %%%%%%%%%%%%
\section{Analysis of Region Three in the Limit $X \to X_c$}
In this section, we describe the behaviour of solutions of the leading order equations for region three in the limit $X \to X_c$, where the velocity $U$ approaches zero. Recall that these leading order equations are 
\begin{align}
 (DU)' &= U Z_b', \label{E:Region3:mass} \\
0 &= D \Delta \rho Z_b' - U^2, \label{E:Region3:mom}\\
(DU\Delta \rho)'  &=\kappa U \Delta T  \label{E:Region3:buoyancy}\\
0&= (1  - 2\Pt -  Z_b)Z_b'U- U\Delta T - DU Z_b',\label{E:Region3:thermal}
\end{align}
and that the flux $Q = DU$ evolves according to
\begin{equation}\label{E:Region3:Q_ODE}
%\left(Q'\right)^3 =\kappa \left(Z_b'\right)^4 \left\{\left(1 - Z_b\right)Q - 2\Pt\left(Q - Q_\in\right) -\left[1 - Z_b(X_p)\right]Q_\in\right\} + \frac{U_\out^3}{Z_b'(X_p)}\left(Z_b'\right)^4.
\frac{\left[Q'(X)\right]^3}{\left[Z_b'(X)\right]^4} = \kappa \left\{ \left[1 - Z_b(X) - 2P_T\right] Q(X) - \left[1 - Z_b(X_p) - 2P_T\right]Q_\text{in}\right\} + \frac{U_\text{out}^3}{Z_b'(X_p)}.
\end{equation}
The point $X_c$ satisfies 
\begin{equation}
\left[1 - Z_b(X_c)\right]Q_c - 2\Pt\left(Q_c - Q_\in\right) -\left[1 - Z_b(X_p)\right]Q_\in +  \frac{U_\out^3}{\kappa Z_b'(X_p)} = 0.
\end{equation}
where $Q_c = Q(X_c)$.

Since the velocity $U \to 0$ as $X = X_c$, we must introduce rescaled variables to reflect a change in asymptotic order; we therefore introduce
%\begin{align}
%U &\sim \kappa^{1/3} Z_b'(X_c)^{2/3} Q_c^{1/3}(X_c - X)^{1/3}, & D &\sim \kappa^{-1/3} Z_b'(X_c)^{-2/3} Q_c^{2/3}(X_c - X)^{-1/3},\label{E:Region3:X_to_Xc1}\\
%\Delta \rho &\sim  \kappa Z_b'(X_c) (X_c - X), & \Delta T &\sim -\kappa^{-1/3} Z_b'(X_c)^{1/3} Q_c^{2/3}(X_c - X)^{-1/3}\label{E:Region3:X_to_Xc2}
%\end{align}
\begin{equation}\label{E:Region3:Rescaling}
X = X_c + \varepsilon \tilde{X}, \quad Q = Q_c + \varepsilon^\gamma \tilde{Q}
\end{equation}
where $\varepsilon \ll 1$ is arbitrary, $\tilde{X} =\order{1}$ is negative, $\tilde{Q} = \order{1}$ and $\gamma >0$ is to be determined. Inserting~\eqref{E:Region3:Rescaling} into~\eqref{E:Region3:Q_ODE} gives
\begin{multline}\label{E:Region3:RescaledODE}
\varepsilon^{3(\gamma - 1)}\frac{\left(\tilde{Q'}\right)^3}{\left[Z_b'(X_c)\right]^4}\left[1 + \order{\varepsilon} \right]= -\varepsilon\lambda\tilde{X}Z_b'(X_c)Q_c + \\ \varepsilon^\gamma \left[1 - 2\Pt - Z_b(X_c)\right] \tilde{Q} + \order{\varepsilon^2, \varepsilon^{\gamma + 1}}.
\end{multline}
A dominant balance is obtained in~\eqref{E:Region3:RescaledODE} by taking $\gamma = 4/3$. After setting $\gamma = 4/3$ in~\eqref{E:Region3:RescaledODE}, using $Q' = U Z_b'$ [from~\eqref{E:Region3:mass}] and undoing the rescaling~\eqref{E:Region3:Rescaling}, we find 
\begin{equation}\label{E:Region3:U_asym}
U \sim \kappa^{1/3} Z_b'(X_c)^{2/3} Q_c^{1/3}(X_c - X)^{1/3} \quad \text{as}~X \to X_c^-.
\end{equation}
From~\eqref{E:Region3:Rescaling}, we have $Q \sim Q_c + \order{\varepsilon^{4/3}}$. Combining this with~\eqref{E:Region3:mass} gives
\begin{equation}\label{E:Region3:D_asym}
D \sim \kappa^{-1/3} Z_b'(X_c)^{-2/3} Q_c^{2/3}(X_c - X)^{-1/3} \quad \text{as}~X \to X_c^-.
\end{equation}
A balance in the momentum equation~\eqref{E:Region3:mom} gives
\begin{equation}\label{E:Region3:drho_asym}
\Delta \rho \sim  \kappa Z_b'(X_c) (X_c - X)\quad \text{as}~X \to X_c^-,
\end{equation}
while a balance in the thermal driving equation~\eqref{E:Region3:thermal} requires
\begin{equation}\label{E:Region3:dT_asym}
 \Delta T \sim -\kappa^{-1/3} Z_b'(X_c)^{1/3} Q_c^{2/3}(X_c - X)^{-1/3} \quad \text{as}~X \to X_c^-.
 \end{equation}

\section{Further Details of Region Four Behaviour}
%scaled variables for this region
\newcommand{\U}{\mathcal{U}}
\newcommand{\D}{\mathcal{D}}
\newcommand{\p}{\Delta \varrho}
\renewcommand{\t}{\Delta \mathcal{T}}
In this section, we provide further details of the behaviour of the leading order equations in region four. Recall that these equations are 
\begin{align}
\dd{(\D\U)}{\chi} &=0, &
\dd{(\D \U^2)}{\chi} &=  \D \p Z_b'(X_c) -\U^2,\label{E:Region4:mass_scaled}\\
\dd{(\D\U\p)}{\chi} &=\kappa  \U\t, &
k_2 \dd{(\D \U \t)}{\chi} &=- \U \t - \D \U Z_b'(X_c).\label{E:Region4:thermal_scaled}
\end{align}
for $ \chi > -\infty$. Equations~\eqref{E:Region4:mass_scaled}--\eqref{E:Region4:thermal_scaled} must satisfy the matching conditions
\begin{align}
\U &\sim \kappa^{1/3}Z_b'(X_c)^{-2/3}Q_c^{1/3} (-\chi)^{1/3}, &  \D &\sim \kappa^{-1/3} Z_b'(X_c)^{-2/3} Q_c^{2/3}(-\chi)^{-1/3}\label{E:Region4:far_field1},\\
\p &\sim - \kappa Z_b'(X_c) \chi, & \t &\sim -\kappa^{-1/3} Z_b'(X_c)^{1/3} Q_c^{2/3}(-\chi)^{-1/3}\label{E:Region4:far_field2}.
\end{align}

We begin by nothing that from the first of~\eqref{E:Region4:mass_scaled}, flux is conserved, i.e.
\begin{equation}\label{E:Region4:Constant_Flux}
    \D \U  = Q_c.
\end{equation}
Also, after inserting~\eqref{E:Region4:Constant_Flux} into the second of~\eqref{E:Region4:thermal_scaled}, we obtain an expression that  can be directly integrated to give
\begin{equation}\label{E:Region4:Linear_p_t}
    \frac{\p}{\kappa} + k_2 \t = -Z_b'(X_c) + C,
\end{equation}
where $C$ is a constant that would be determined by analysing the higher order contributions to $\Delta \rho$ as $X \to X_c$. This higher order analysis is beyond the scope of this paper.

The relationships~\eqref{E:Region4:Constant_Flux} and~\eqref{E:Region4:Linear_p_t} allow us to reduce~\eqref{E:Region4:mass_scaled}--\eqref{E:Region4:thermal_scaled} to a system of two equations, which can be rescaled so that only a single parameter, $k_2$, enters:
\begin{equation}\label{E:Region4:rescaled}
\tilde{\U} \dd{\tilde{\U}}{\tilde{\chi}} = \tilde{\p} - \tilde{\U}^3, \qquad 
k_2  \dd{\tilde{\p}}{\tilde{\chi}} = -\tilde{\U}\left[\tilde{p} + \tilde{\chi}\right],
\end{equation}
where
\begin{align}
\tilde{\chi} &= \frac{\kappa^{1/4}Z_b'(X_c)^{1/2}}{Q_c^{1/2}}\left(\chi - \frac{C}{\kappa Z_b'(X_c)}\right), \label{E:Region4:RescalingChi} \\ 
\tilde{\U} &= \frac{1}{\kappa^{1/4}Z_b'(X_c)^{1/2}Q_c^{1/2}}\U, \label{E:Region4:RescalingU}\\ 
\tilde{\p}&= \frac{1}{\kappa^{3/4}Z_b'(X_c)^{ 1/2}Q_c^{1/2}}\p. \label{E:Region4:RescalingP}
\end{align} 
In terms of these rescaled variables, the matching conditions~\eqref{E:Region4:far_field1}--\eqref{E:Region4:far_field2} read
\begin{equation}\label{E:Region4:Rescaled_farfield}
\tilde{\p} \sim -\tilde{\chi},~ \tilde{\U}\sim (-\tilde{\chi})^{1/3}\quad \text{as}~\tilde{\chi}\to -\infty.
\end{equation}

\begin{figure}
\centering
\includegraphics[width = 0.95\textwidth]{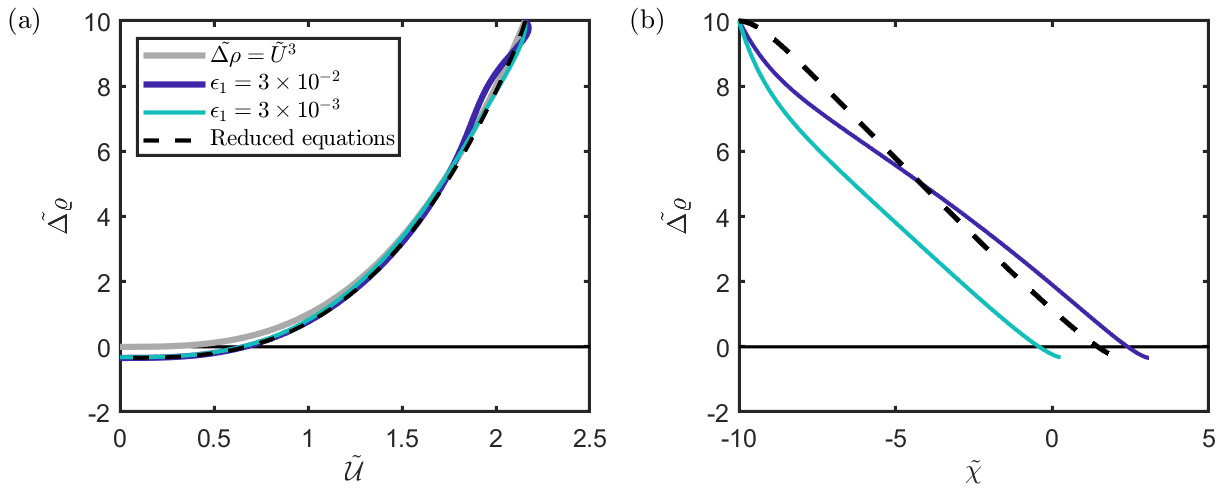}
\caption{Numerical solutions of original model equations [(2.30)--(2.33) in the main text] rescaled according to~(3.31)--(3.32) and with $\epsone = 3\times 10^{-2}$ (i.e. as in table 2 of the main text, purple curves) and with $\epsone = 3\times 10^{-3}$ (cyan curves) which are shown in (a) $(\tilde{\U}, \tilde{\p})$ space and (b) $(\tilde{\chi}, \tilde{\p})$ space [$\tilde{\chi}$, $\tilde{\U}$, and $\tilde{\p}$ are a spatial variable, plume velocity, and buoyancy deficit, respectively, and are defined in~\eqref{E:Region4:RescalingChi}--\eqref{E:Region4:RescalingP}]. The black dashed curve indicates the numerical solution of the reduced equations~\eqref{E:Region4:rescaled}, and the grey curve indicates the nullcline $\tilde{\U}^3 = \tilde{\p}$. The solid black line indicates $\tilde{p} = 0$. Each of the solutions here uses a linear draft, $Z_b(X) = X$, with $Q_c = 0.5, X_c = 0.5$ and $\kappa, k_i$ according to the values in table 2 of the main text. The matching conditions~\eqref{E:Region4:far_field1}--\eqref{E:Region4:far_field2} are applied at $\tilde{\chi}_0 = 10$ in both cases, i.e. the solution domain shown is $-\tilde{\chi}_0 < \tilde{\chi} < \tilde{\chi}_t$, where $\tilde{\chi}_t$ is the value of $\tilde{\chi}$ at which the plume velocity reaches zero.}\label{fig:Region4}
\end{figure}

The system~\eqref{E:Region4:rescaled}--\eqref{E:Region4:Rescaled_farfield} must be solved numerically. In figure~\ref{fig:Region4}, we present a comparison between numerical solutions of equations~\eqref{E:Region4:rescaled} and the solutions of the full equations [model equations~(2.30)--(2.33) in the main text rescaled according to~(3.31)--(3.32)]. We see good agreement, with solutions predominantly following the nullcline $\tilde{\p} = \tilde{\U}^3$, before deviating when the rescaled buoyancy deficit $\tilde{\p}$ approaches zero. Eventually, the buoyancy deficit goes negative, indicating that the plume has become negatively buoyant; the plume's upward motion continues for a short distance owing to inertia, before ultimately reaching $\tilde{\U} = 0$, where it terminates. For the purpose of constructing an approximation to the melt rate, it is worth noting that the rescaled buoyancy is approximately linear (figure~\ref{fig:Region4}b); using this, the nullcline solution for $\tilde{\U}$, which holds nearly everywhere in this region, can be approximated by $\tilde{\U} = \tilde{\chi}^{1/3}$. After undoing the rescaling~\eqref{E:Region4:RescalingChi}--\eqref{E:Region4:RescalingP}, this approximate solution reduces to equation (3.37) in the main text. 

%%%%%%%%%%%%%%%%%%%% Melt Rate in Exception Cases %%%%%%%%
\section{Melt Rate Construction in Special Cases}
In this section, we explicity set out our constructed melt rate profiles in the special cases described in \S4 of the main text.

The first special case arises when $\Delta \rho_{\in} - 2 \Pb Z_b'(X_p) < 0$. Our analysis suggests that, in this case, the plume will intrude into the ambient within the pycnocline. Our constructed melt rate takes a linear interpolation across the pycnocline as in the approximation presented in the main text [equation (4.12) therein] that is modified to include zero speed and thermal driving upon exiting:
\begin{equation}\label{E:MeltRate1}
M_{p} = \begin{cases} 
M_{p,1~} \text{~\quad [equation~(4.1)]}  & 0 < X < X_p - N_l \lt,\\
M_{p,2~} \text{~\quad [equation~(4.4)]} & X_p - N_l \lt < X <X_{\text{sep}} ,\\
0  &  X > X_{\text{sep}},\\
\end{cases}
\end{equation}
where $X_{\text{sep}}$ is the value of $X$ at which $M_{p,2} = 0$. \blue{We note that the formulation~\eqref{E:MeltRate1} does not account for the subregion within the pycnocline in which the buoyancy deficit ceases to be $\order{1}$ and the plume velocity reaches zero; this is justified on account of this region having a lengthscale which is $\order{\epsone \delta}$, meaning that it is unimportant on the lengthscale of the entire shelf.}

The second exception case occurs when no physically relevant solution of~(4.7), which describes the `cross-over' point $X^*$ must satisfy, exists. In this case,  the scaled melt rate takes values
\begin{equation}\label{E:MeltRate:AllRegions_noroot}
M_{p} = \begin{cases} 
M_{p,1~} \text{~\quad [equation~(4.1)]}  & 0 < X < X_p - N_l \lt,\\
M_{p,2~} \text{~\quad [equation~(4.4)]} & X_p - N_l \lt < X < X_p + N_l \lt,\\
M_{p,3l~} \text{\quad [equation~(4.8)]} &  X > X_p + N_l \lt.
\end{cases}
\end{equation}
Finally, if the computed termination point $X_c$ does not satisfy $X_c > X^*$, we take $M_p$ as in~\eqref{E:MeltRate:AllRegions_noroot}.